\documentclass[manuscript,screen,nonacm]{acmart}
\AtBeginDocument{%
  \providecommand\BibTeX{{%
    Bib\TeX}}}

\usepackage{multirow}
\usepackage{subfig}
\usepackage{bbding}
\usepackage{float}
\usepackage[table,xcdraw]{xcolor}
\usepackage{booktabs} 

\usepackage{soul} 

\usepackage{todonotes}
\def\BibTeX{{\rm B\kern-.05em{\sc i\kern-.025em b}\kern-.08em
    T\kern-.1667em\lower.7ex\hbox{E}\kern-.125emX}}

\newcommand{\DrawPercentageBar}[1]{%
  \begin{tikzpicture}
    \fill[color=black]   (0.0 , 0.0) rectangle (#1*3ex , 1.5ex );
    \fill[color=gray] (#1*3ex  , 0.0) rectangle (3.0ex, 1.5ex);
  \end{tikzpicture}%
}

\newcommand{\toolname}[1]{\emph{VulnGym}}

\setcopyright{none}

\begin{document}

\title{VulnGym: Evaluating Vulnerability Management Strategies against Advanced Persistent Threats}


\emergencystretch=2em

\author{Sofia Della Penna}
\email{sofia.dellapenna@unina.it}
\author{Lorenzo Parracino}
\email{lorenzo.parracino@unina.it}
\author{Luciano Pianese}
\email{luciano.pianese@unina.it}
\author{Vittorio Orbinato}
\email{vittorio.orbinato@unina.it}
\affiliation{%
  \institution{Università degli Studi di Napoli Federico II}
  \city{Naples}
  \country{Italy}}

\author{Roberto Natella}
\email{roberto.natella@gssi.it}
\affiliation{%
  \institution{Gran Sasso Science Institute}
  \city{L'Aquila}
  \country{Italy}
}

\renewcommand{\shortauthors}{Della Penna et al.}


\begin{abstract}
Enterprise networks are continuously targeted by Advanced Persistent Threats (APTs), attack campaigns exploiting software vulnerabilities to compromise critical assets over time. As disclosed vulnerabilities grow, resource-constrained organizations must prioritize which ones to patch. Existing prioritization standards score vulnerabilities individually and cannot capture how a patching policy performs against an adversary that progresses through the network over time.
Previous tools have simulated attack campaigns through Reinforcement Learning (RL), but either omit vulnerability management, leaving the attacker unopposed, or rely on synthetic networks disconnected from real threat data, and so cannot assess how a policy would fare against a realistic adversary.
To fill this gap, we propose \toolname{}, a simulation tool to evaluate vulnerability management policies. \toolname{} simulates an RL-trained attacker, calibrated on real APT profiles, against a defender executing a configurable patching policy over a network with real Common Vulnerabilities and Exposures (CVEs). Both agents act on a shared, evolving network representation, so the attacker's progress is directly shaped by the defender's patching activity, allowing a given policy to be stress-tested against a realistic attack campaign.
Experiments based on real-world vulnerabilities and two APTs show that vulnerability management must be tailored to organizational context, adversarial behavior, network topology, and asset criticality.
\end{abstract}



\begin{abstract}
Enterprise networks are continuously targeted by Advanced Persistent Threats (APTs), attack campaigns exploiting software vulnerabilities to compromise critical assets over time. As disclosed vulnerabilities grow, resource-constrained organizations must prioritize which ones to patch. Existing prioritization standards score vulnerabilities individually and cannot capture how a patching policy performs against an adversary that progresses through the network over time.
Previous tools have simulated attack campaigns through Reinforcement Learning (RL), but either omit vulnerability management, leaving the attacker unopposed, or rely on synthetic networks disconnected from real threat data, and so cannot assess how a policy would fare against a realistic adversary.
To fill this gap, we propose \toolname{}, a simulation tool to evaluate vulnerability management policies. \toolname{} simulates an RL-trained attacker, calibrated on real APT profiles, against a defender executing a configurable patching policy over a network with real Common Vulnerabilities and Exposures (CVEs). Both agents act on a shared, evolving network representation, so the attacker's progress is directly shaped by the defender's patching activity, allowing a given policy to be stress-tested against a realistic attack campaign.
Experiments based on real-world vulnerabilities and two APTs show that vulnerability management must be tailored to organizational context, adversarial behavior, network topology, and asset criticality.
\end{abstract}
\maketitle
\section{Introduction}
\label{sec:introduction}

In the current cybersecurity landscape, organizations are increasingly confronted with sophisticated attacks, known as Advanced Persistent Threats (APTs)  \cite{microsoft_apt,google_apt}.
APTs apply a variety of techniques to stealthily propagate across computer networks, to evade detection, to maintain persistence over extended periods, and ultimately to disrupt services and exfiltrate sensitive data \cite{mitre_attack,lockheed_cyber_kill_chain,Singh2019}.
In particular, APT attacks are enabled by \emph{software vulnerabilities} in the network, which can be exploited to obtain initial access, gain privileges, and move laterally across the network. 
Notable examples of high-impact vulnerabilities include the Microsoft Server Message Block (SMB) vulnerability exploited by the \textit{WannaCry} ransomware through the \textit{EternalBlue} exploit \cite{NAO2017}, the Apache Struts vulnerability exploited in the large data breach from the \textit{Equifax} agency \cite{wang2018cybersecurity}, and the \textit{Log4Shell} vulnerability in the popular Log4j open-source library \cite{feng2022defense}. 
Therefore, software vulnerabilities represent a critical risk faced by organizations today.


Despite the continuous release of security patches for newly discovered vulnerabilities, a substantial proportion of cyberattacks still exploit known vulnerabilities for which patches have long been available but not applied \cite{recordedfuture2024vulns, csw2024ransomware,bitsight_critical_updates}. 
%
%
This problem is caused by the overwhelming volume of reported vulnerabilities that IT teams must manage: the number of vulnerabilities reported in the National Vulnerability Database (NVD) was close to $50,000$ new entries in 2025 \cite{vulnerability_stats,cveicu}. As a matter of fact, organizations have a limited capacity to deploy patches \cite{empirical2026capacity}, since vulnerability management requires significant resources in terms of time and personnel, and since patches can disrupt business operations, e.g., due to human errors and broken dependencies \cite{nurse2025patch,NIST80040r4}. 
This trend is expected to be exacerbated by the emergence of LLM-based tools capable of autonomously discovering and exploiting software vulnerabilities, a capability recently demonstrated at scale by frontier AI models, which is likely to substantially increase the volume of vulnerability reports organizations must handle \cite{anthropic2026glasswing}.

To keep up with the pace of vulnerabilities, several standards and metrics have been developed to guide vulnerability management, by prioritizing efforts on the most critical vulnerabilities \cite{NIST80040r4,NCSC2023,CVEorg,EPSS,CVSS,CISAKEV,CISA_SSV}, such as vulnerabilities that are easier to exploit or that have already been exploited in recent attacks. However, these guidelines only analyze vulnerabilities individually, and do not consider how vulnerabilities can be leveraged in the wider context of a complete attack campaign. 
As a matter of fact, attack campaigns from APTs combine vulnerability exploitation with several other post-exploitation techniques, such as lateral movement, credential stealing, and persistence. Moreover, multiple vulnerabilities can be exploited by the same attack campaign. Therefore, the prioritization of vulnerability management should take into account multiple factors, including the attack patterns of APTs, and the topology and assets of the victim network.

%

In this paper, we present \toolname{}, a novel simulation tool\footnote{Available at: \url{https://github.com/dessertlab/vulnGym}} for the quantitative analysis of vulnerability management strategies. 
\toolname{} models the dynamics of complex attack campaigns in IT networks, by taking into account that: (i) new vulnerabilities are being discovered over time; (ii) IT teams continuously scan for vulnerabilities and deploy patches to fix them, with a limited budget; (iii) adversaries can exploit vulnerabilities towards compromising sensitive assets. 
In particular, the tool uses Reinforcement Learning (RL) to train an adversarial agent able to orchestrate multiple attack techniques and to maximize impact on assets. 
Moreover, the tool enables the modeling of vulnerability management policies, such as policies based on severity scores (e.g., CVSS), and policies that prioritize specific assets and network segments. The analysis brings insights on how effective such policies are at mitigating attack campaigns as a whole, providing valuable feedback for vulnerability management.

The paper includes an analysis of vulnerability management policies with respect to real vulnerabilities from the NVD \cite{nist_nvd}. 
The analysis shows the shortcomings of vulnerability management based on \emph{severity scores} of individual vulnerabilities, and the benefits of holistic policies that account for \emph{asset criticality}. This result motivates the need to consider more contextual information in defining prioritization policies, including the expected adversarial behaviors, the budget for vulnerability management, and the topology and assets of the network.



    
    
    

The remainder of the paper is organized as follows. Section~\ref{sec:background} provides background on vulnerability management frameworks and Reinforcement Learning. Section~\ref{sec:related_work} reviews the related work. Section~\ref{sec:methodology} presents the proposed simulation approach. Sections~\ref{sec:experimental_design} and~\ref{sec:evaluation} describe the experimental design and evaluate the proposed vulnerability management policies. Section~\ref{sec:threats} discusses the threats to validity. Finally, Section~\ref{sec:conclusion} concludes the paper.
\section{Background}
\label{sec:background}
\subsection{Vulnerability Management Frameworks}
Vulnerability management is a cornerstone of cybersecurity. It is based on frameworks for \textit{Risk Assessment}, \textit{Vulnerability Identification and Prioritization}, and \textit{Severity Scoring}.

Risk management frameworks provide structured approaches for securing information systems \cite{NIST_CSWP_29}. Among these, NIST Special Publication 800-40 \cite{NIST80040r4} offers recommendations for operationalizing vulnerability management. It advocates for patch prioritization based on cybersecurity impact, integration with change management processes, secure acquisition and validation of patches, and thorough testing in staging environments prior to deployment.

For vulnerability identification and classification, the Common Vulnerabilities and Exposures (CVE) registry \cite{CVEorg} serves as the \emph{de facto} standard. Each CVE entry receives a unique identifier and description from designated CVE Numbering Authorities (CNAs). The Known Exploited Vulnerabilities (KEV) Catalog \cite{CISAKEV} by CISA extends the CVE framework by tracking vulnerabilities that were exploited in the wild.

Severity scoring and exploit likelihood are commonly estimated using the Common Vulnerability Scoring System (CVSS) \cite{CVSS}, which quantifies vulnerability impact across multiple technical dimensions. The Exploit Prediction Scoring System (EPSS) \cite{EPSS} complements CVSS by leveraging machine learning models trained on threat intelligence data to estimate the probability that a vulnerability will be exploited in the wild.

Despite their widespread adoption, these scoring systems have repeatedly been questioned for their ability to reflect real exploitation risk. Younis et al. \cite{younis2015comparing} compared CVSS Base metrics and Microsoft's rating system over 813 vulnerabilities, revealing high false-positive rates and a strong dependency on software-specific factors. Several subsequent studies confirmed that many low-scoring vulnerabilities are still exploited in the wild \cite{garrity_insights_2023, islam_cvss_deception_2024, redhat_cvss_not_risk_2019, cisco_p2p_vol9_2023}. For instance, ransomware campaigns have repeatedly exploited CVEs classified as medium/low severity \cite{edwards_bitsight_kev_2024}, revealing that CVSS lacks contextual awareness \cite{fernao_cvss_misconceptions_2024, shick_towards_cvss_2018}. Similarly, EPSS, despite improving on CVSS through machine learning, remains an opaque approach unable to replace systematic risk analysis \cite{sei2022epss}. Taken together, these findings show that individually-scored vulnerabilities are a poor proxy for the risk posed by a real attack campaign, which motivates research in the field as discussed in Section~\ref{sec:related_work}.

\subsection{Reinforcement Learning}
We adopt \textit{Reinforcement Learning} (RL) \cite{ADAWADKAR2022} to create automated agents that emulate attackers. 
RL is a machine learning paradigm in which a decision-making agent interacts actively with an uncertain environment. 
The key components of a RL system are as follows \cite{ADAWADKAR2022}:
\begin{itemize}
    \item \textit{Agent}: seeks goals by interacting with the environment.
    \item \textit{Environment}: everything the agent interacts with. The agent reads the state of the environment to choose actions.
    \item \textit{Policy}: a map between the states in the environment and the actions that are taken from that state.
    \item \textit{Reward Signal}: defines the goal of the problem. At each time step, RL assesses the current state and action of the agent, and sends a reward value to the agent.
    \item \textit{Value Function}: is defined for a state. When the agent starts from a state and accumulates a total reward over time, it is known as the value of the state.
\end{itemize}

\textit{Q-Learning} (QL) \cite{watkins1992q} is a model-free RL algorithm, based on the \textit{Q-function} \(Q(s, a)\), which estimates the expected reward of taking action \(a\) in state \(s\) to guide the agent and is updated iteratively according to Eq.~\ref{eq:qlearning_update}:
\begin{equation}
\scalebox{0.83}{$Q(s_t, a_t) \leftarrow Q(s_t, a_t) + \alpha \big[ R_{t+1} + \gamma \max_a Q(s_{t+1}, a) - Q(s_t, a_t) \big]$}
\label{eq:qlearning_update}
\end{equation}
where:
\begin{itemize}
    \item \( Q(s_t, a_t) \) is the current estimate of the \textit{Q-value} for state \( s_t \) and action \( a_t \).
    \item \( R_{t+1} \) is the immediate reward received after taking action \( a_t \) in state \( s_t \).
    \item \( \gamma \) is the discount factor, a parameter between $0$ and $1$ that represents the importance of future rewards relative to immediate rewards.
    \item \( \alpha \) is the learning rate, which controls the magnitude of updates to the \textit{Q-values} based on new experiences.
    \item \( \max_a Q(s_{t+1}, a) \) is the maximum \textit{Q-value} over all possible actions in the next state \( s_{t+1} \), representing the optimal future value.
\end{itemize}

\textit{Deep Q-Learning} or \textit{Deep Q-Network} (DQN) \cite{fan2020theoretical, mnih2013playing} is an extension of QL, where the \textit{Q-function} is approximated by a Deep Neural Network \( Q(s, a; \theta) \), where \( \theta \) denotes the network weights. It implements experience replay at each time step \( t \), by storing the agent’s experiences \( (s_t, a_t, r_t, s_{t+1}) \) in a replay buffer. The DQN update at iteration \( i \) is performed by minimizing the loss function in Eq.~\ref{eq:dqn_loss}:
\begin{equation}
L_i(\theta_i) = \mathbb{E}_{(s_t, a_t, r_t, s_{t+1})} \left[ \left( y_i - Q(s_t, a_t; \theta_i) \right)^2 \right]
\label{eq:dqn_loss}
\end{equation}
\begin{equation}
y_i = r_t + \gamma \max_{a'} Q(s_{t+1}, a_{t+1}; \theta_i^{-})
\label{eq:dqn_target}
\end{equation}
where:
\begin{itemize}
    \item \( \mathbb{E}_{(s_t, a_t, r_t, s_{t+1})} \) is the average on replay buffer samples.
    \item \( Q(s_t, a_t; \theta_i) \) is the predicted \textit{Q-value}.
    \item \( Q(s_{t+1}, a_{t+1}; \theta_i^{-}) \) is the target \textit{Q-value} for next state.
\end{itemize}
\section{Related Work}
\label{sec:related_work}

Several tools have been proposed to simulate attack campaigns and vulnerability management processes, ranging from optimization-based patch prioritization to graph-based attack analysis and, more recently, RL-based simulation environments.
 
Among optimization-based tools, Farris et al. \cite{farris2018vulcon} proposed \textit{VULCON}, a vulnerability management framework that uses a mixed-integer goal-programming model to select which vulnerabilities to patch at each remediation cycle, jointly minimizing vulnerability exposure and remediation delay under a personnel-hour budget. However, the framework focuses exclusively on the defender and does not model attacker behavior or vulnerability exploitation. \toolname{} extends this perspective by pairing a budget-constrained defender with an RL-trained attacker calibrated on real APT attack profiles, enabling defensive policies to be evaluated against realistic attack campaigns.

Attack graphs have also been proposed to analyze the impact of vulnerabilities on network security. \textit{MulVAL} \cite{ou2005mulval} is a logic-based framework that models vulnerabilities, host configurations, and network connectivity using Datalog, to enumerate all potential attack paths across the network graph. Similarly, the \textit{Topological Vulnerability Analysis} (TVA) \cite{singhal2017security} and \textit{Bayesian attack graph models} \cite{munoz2017exact} apply this approach for risk quantification. These tools analyze how vulnerabilities can be combined in complex attack scenarios, but they rely on static snapshots of the network, and do not jointly model the temporal dynamics of defenders that perform vulnerability management, of new vulnerabilities discovered over time, and of the adaptive behavior of adversaries. In contrast, \toolname{} focuses on the evaluation of vulnerability management in a dynamic setting, explicitly modeling the interaction between defenders and attackers over time, and leverages RL to model sophisticated APT behaviors.

Closer to our approach, RL has been used to train autonomous attacker agents and, in some cases, defender agents in simulated network environments.
Becker et al. \cite{becker2024evaluationreinforcementlearningautonomous} extended \textit{NASim} \cite{schwartz2019autonomouspenetrationtestingusing}, a lightweight Gymnasium-based network attack simulator, to train RL agents for penetration-testing scenarios involving exploitation, credential theft, and wiretapping, comparing their performance against a rule-based penetration-testing baseline. However, their evaluation is limited to small network topologies of four to five hosts and excludes active defense and vulnerability remediation, allowing the attacker to operate without opposition.
\textit{CyberBattleSim} \cite{cyberbattlesim} is an OpenAI Gym environment developed by Microsoft that models an enterprise network as a graph of abstract nodes, in which a trainable attacker exploits local and remote vulnerabilities and performs lateral movement to compromise network assets. However, the simulator relies on synthetic, abstract vulnerabilities rather than real CVEs and does not model the vulnerability management lifecycle.
In contrast to \textit{NASim} and \textit{CyberBattleSim}, \toolname{} grounds both the network and the vulnerabilities in real-world data from the NVD, models the defender through configurable vulnerability management policies operating under a limited patching budget, and trains the attacker on behavior profiles derived from real APT threat intelligence, allowing a direct, quantitative evaluation of vulnerability prioritization strategies against realistic adversarial campaigns.
\section{Methodology}
\label{sec:methodology}

The basic idea of our approach is to simulate the interactions between an \textit{attacker} and a \textit{defender} (\emph{agents}) within a network with vulnerable assets, in order to analyze the impact of vulnerability management policies against attack campaigns.

\begin{figure}[H]
    \centering
    \includegraphics[width=0.7\linewidth]{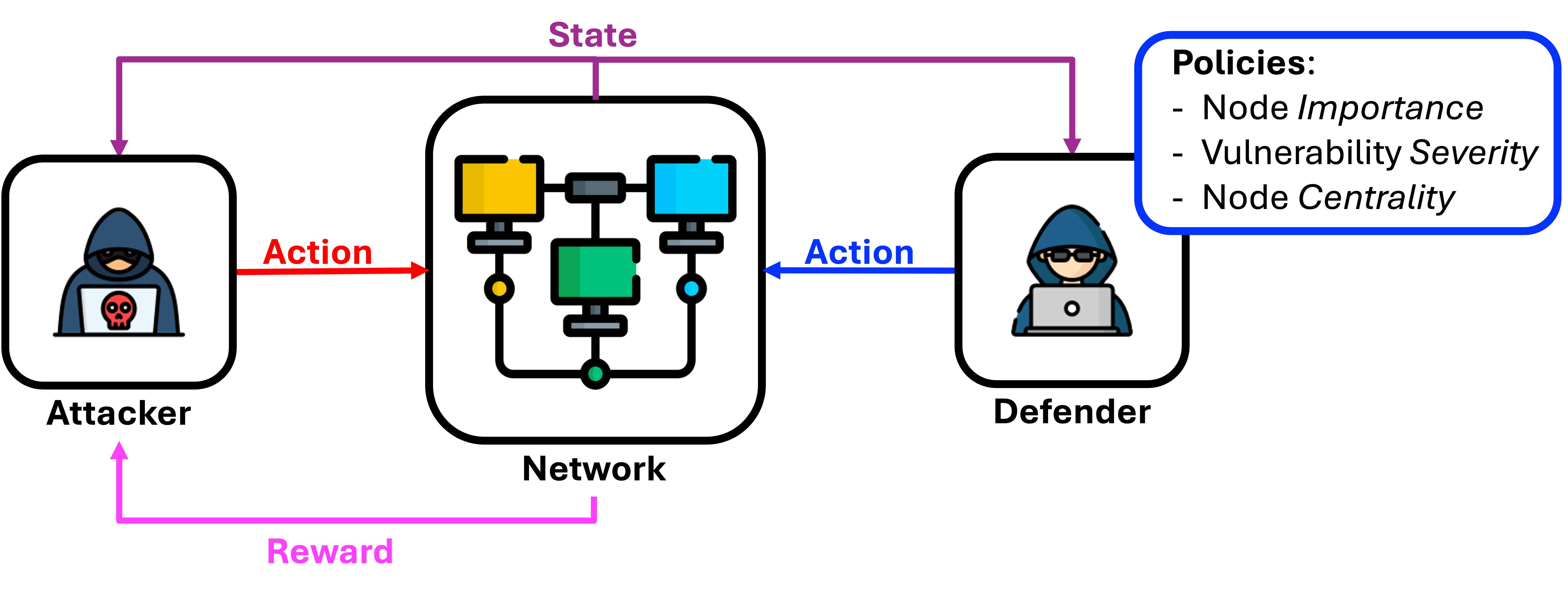}
    \caption{Architecture of the simulated environment.}
    \label{fig:architecture_schema}
\end{figure}

Figure \ref{fig:architecture_schema} summarizes the elements of the simulation. The agents perform \emph{actions} on the nodes of the simulated network. 
Actions are selected based on the current \emph{state} of the network (e.g., nodes that have already been compromised) and modify the state of the affected nodes. After executing an action, the attacker agent receives a \emph{reward}, a numerical representation of its outcome that can be either positive or negative. The reward is used to train it via RL.
Through this process, the agent learns to orchestrate a sequence of attack techniques that maximize the  overall impact of the attack on the network.
In parallel, the defender agent follows a policy to mitigate vulnerabilities across the network, representing a vulnerability management strategy to be evaluated (e.g., based on exploitability scores).

\begin{figure}[ht]
    \centering
    \includegraphics[width=0.5\linewidth]{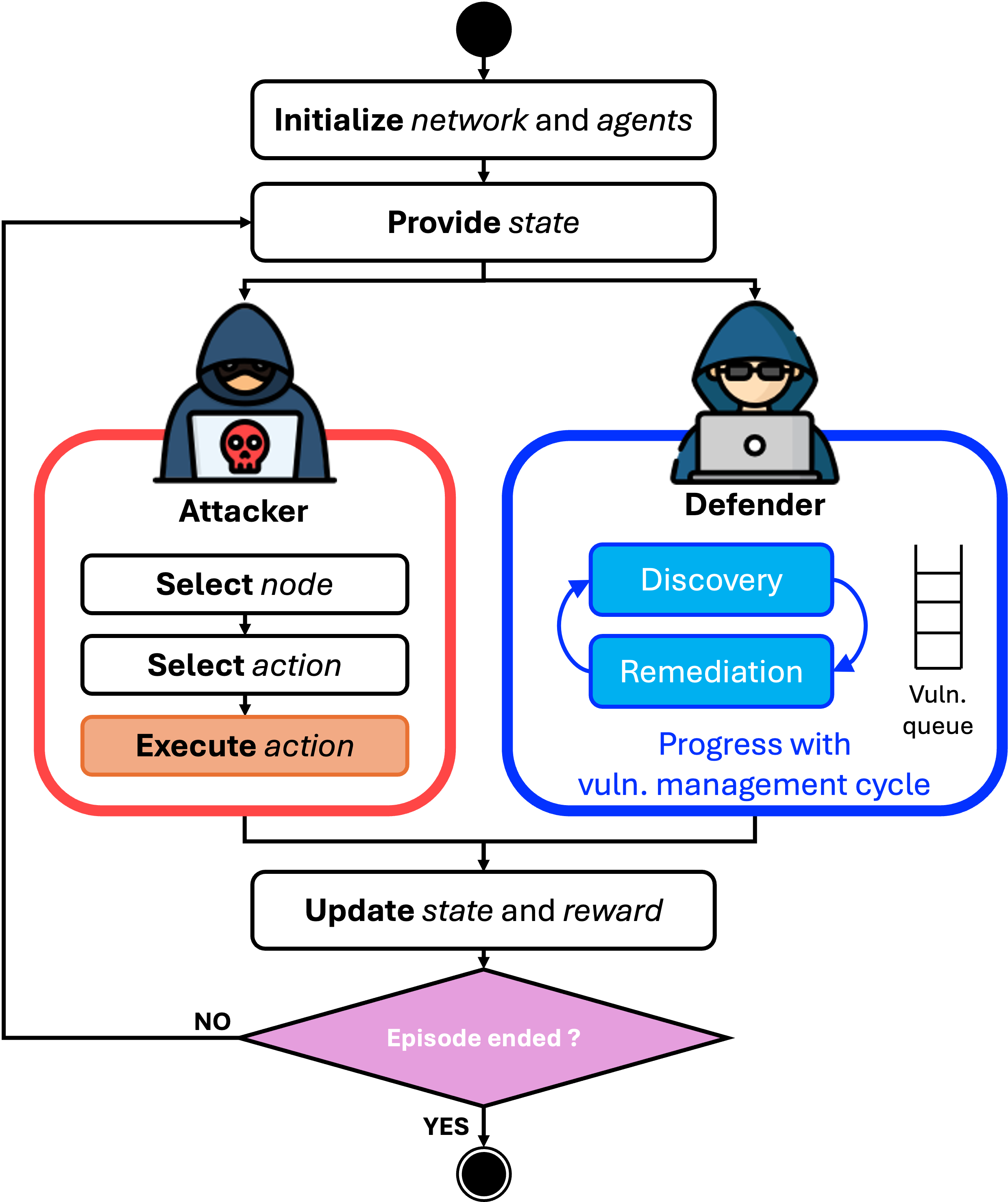}
    \caption{Execution flow of an episode.}
    \label{fig:execution_flow}
\end{figure}

Figure~\ref{fig:execution_flow} illustrates the execution flow of a single \emph{episode}, representing a simulated time period (e.g., one year). At the beginning of the episode, the network environment and the agents are initialized. Both agents then interact with the environment: the attacker selects a target node and an action, then executes it; the defender progresses through a vulnerability management cycle, alternating between discovery and remediation phases while maintaining a vulnerability queue. Multiple episodes are simulated to account for the stochastic nature of the simulation, including probabilistic agent behavior and uncertain action outcomes.

\subsection{Network}

The simulated network is the environment in which agents operate and interact dynamically, and is designed to emulate realistic enterprise conditions in terms of both topology and vulnerabilities. 
The network is modeled as an undirected graph, where nodes represent potentially vulnerable assets, and edges define the connections between them, shaping the possible paths for attacker movement.

The network is logically divided into zones, which are typical of IT enterprise networks \cite{bansal2020zoning}. 
The network is fully customizable and supports the definition of multiple zones (e.g., DMZ, Database) as well as intra-zone and inter-zone connections. Examples of network topologies are provided in the experimental evaluation of the paper (Section~\ref{sec:experimental_design}). 

Each node represents a network resource and is characterized by the attributes listed in Table~\ref{tab:node_attributes}.
%
%
Asset criticality is quantified through two dedicated attributes: \textit{Importance}, reflecting the operational value of the node within the network, and \textit{Centrality}, a graph-theoretic measure of how pivotal the node is with respect to the overall network  topology.
Each node is also associated with one or more operating system and application products, selected from those commonly used in enterprise environments. Examples include OS products such as Microsoft Windows, Microsoft Windows Server, and Red Hat Linux, and applications from Adobe, Oracle, and Microsoft.
Moreover, we define the \textit{Attacker Access} attribute to track whether the node has been successfully compromised.

\begin{table}[htbp]
\centering
\caption{Node attributes.}
\label{tab:node_attributes}
\begin{tabular}{|l|l|}
\hline
\multicolumn{1}{|c|}{\textbf{Name}} & \multicolumn{1}{c|}{\textbf{Description}} \\ \hline
ID                                  & Unique identifier of the node.                                \\ \hline
ZONE                                & Network zone to which the node belongs.                              \\ \hline
IMPORTANCE                          & Importance level of the node within the network.             \\ \hline
CENTRALITY                          & Node's structural relevance in the network.             \\ \hline
PRODUCTS                             & List of software products installed on the node.                          \\ \hline
VULNERABILITIES                     & List of CVEs affecting the node.                   \\ \hline
STATE                               & Current operational state of the node.                             \\ \hline
ATTACKER ACCESS                   & Whether the attacker has gained access to the node.                           \\ \hline
\end{tabular}
\end{table}

\toolname{} allows the user to configure which vulnerabilities to introduce in the simulation, and when to introduce them. This information can be obtained from public sources, such as the NVD \cite{nist_nvd}. 
Table~\ref{tab:vuln_attributes} lists the attributes that can be configured for each vulnerability.
Among these, \textit{Vulnerability Type} is assigned by evaluating the impact of the vulnerability on the CIA triad as reported in the CVSS vector string. If the vulnerability reports a non-zero impact on Confidentiality or Integrity, it is labelled as \textit{Remote Control}, as it may allow an attacker to access or manipulate data on the node. If the impact on Availability is non-zero, it is labelled as \textit{Denial of Service}, reflecting the attacker's ability to disrupt the normal operation of the node. A vulnerability can be assigned both types when it affects multiple dimensions of the triad simultaneously. 
The simulation models the progressive discovery and exploitation of vulnerabilities within the network. Initially, each vulnerability is considered undisclosed, and it is gradually enabled according to its actual publication date. When a vulnerability is disclosed, it becomes active on all nodes that host a product affected by that vulnerability.
Attackers must adapt their behavior as new vulnerabilities emerge, while defenders must decide how to prioritize their mitigation efforts. 
%
%
\begin{table}[H]
\centering
\caption{Vulnerability attributes.}
\label{tab:vuln_attributes}
\begin{tabular}{|l|l|}
\hline
\multicolumn{1}{|c|}{\textbf{Name}} & \multicolumn{1}{c|}{\textbf{Description}}     \\ \hline
CVE ID                              & Unique identifier of the vulnerability.           \\ \hline
CVSS SCORE                          & Severity score assigned to the vulnerability according to the CVSS.               \\ \hline
VULNERABILITY TYPE                  & Remote Control and/or Denial of Service. \\ \hline
RELEASE TIME                        & Date on which the vulnerability was publicly disclosed.              \\ \hline
ATTACK COMPLEXITY                   & Level of effort required for a successful exploitation.         \\ \hline
\end{tabular}
\end{table}

\subsection{Attacker}

The tool simulates an \textit{attacker} agent based on DQN, which follows the policy learned through RL. 

\begin{figure*}[ht]
    \centering
    \includegraphics[width=\linewidth]{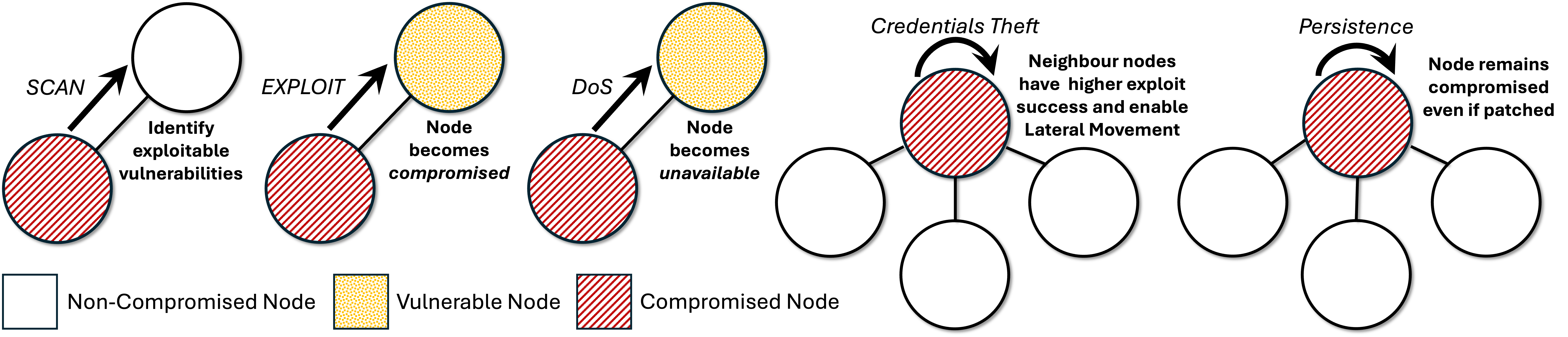}
    \caption{Propagation of the attack across the network.}
    \label{fig:simulation}
\end{figure*}

The attacker operates over the action set listed in Table~\ref{tab:attackers_actions}. This set is designed to reflect the intrusion process as described by the Cyber Kill Chain \cite{lockheed_cyber_kill_chain}, and it is aligned with the tactics and techniques defined in the MITRE ATT\&CK framework \cite{mitre_attack}. 
Figure~\ref{fig:simulation} provides a graphical overview of the attacker’s actions, showing the targets and the effects on the simulation.

\begin{table*}[ht]
\centering
\caption{Attacker's actions.}
\label{tab:attackers_actions}
\resizebox{\textwidth}{!}{%
\begin{tabular}{|l|l|l|l|}
\hline
\multicolumn{1}{|c|}{\textbf{Action}} & \multicolumn{1}{c|}{\textbf{Target}} & \multicolumn{1}{c|}{\textbf{Description}}                                                                                            & \multicolumn{1}{c|}{\textbf{Post-Condition}}                                                                                                            \\ \hline
SCAN &
  \textit{non-compromised node} &
  \begin{tabular}[c]{@{}l@{}}The attacker scans the node for exploitable vulnerabilities.\end{tabular} &
  \begin{tabular}[c]{@{}l@{}}The attacker knows the node's exploitable vulnerabilities.\end{tabular} \\ \hline
EXPLOIT &
  \textit{vulnerable node} &
  \begin{tabular}[c]{@{}l@{}}The attacker launches a remote-control attack on the node.\end{tabular} &
  The attacker gains access to the node. \\ \hline
DOS &
  \textit{vulnerable node} &
  \begin{tabular}[c]{@{}l@{}}The attacker launches a DoS attack on the node.\end{tabular} &
  The node is unavailable. \\ \hline
CREDENTIALS THEFT &
  \textit{compromised node} &
  The attacker steals credentials from the node. &
  \begin{tabular}[c]{@{}l@{}}Increases exploitation success and enable \textit{Lateral} \\ \textit{Movement} on neighbouring nodes.\end{tabular} \\ \hline
PERSISTENCE                           & \textit{compromised node}            & The attacker establishes persistence on the node.                                                                                    & \begin{tabular}[c]{@{}l@{}}If the defender patches the node, the attacker still retains \\ access to it.\end{tabular}                                   \\ \hline
PRIVILEGE ESCALATION                  & \textit{compromised node}            & The attacker escalates privileges on the node.                                                                                       & \begin{tabular}[c]{@{}l@{}}Increases the probability of successfully stealing \\ credentials and achieving persistence on the node.\end{tabular}                        \\ \hline
DATA EXFILTRATION                     & \textit{compromised node}            & The attacker exfiltrates data from the node.                                                                                         & \begin{tabular}[c]{@{}l@{}}The attacker is rewarded when targeting database nodes \\ with sensitive assets (\textit{Database} layer); otherwise, no effect.\end{tabular} \\ \hline
WIPER                                 & \textit{compromised node}            & The attacker deletes or corrupts the node’s data.                                                                                    & \begin{tabular}[c]{@{}l@{}}The attacker is rewarded when targeting database nodes \\ with sensitive assets (\textit{Database} layer); otherwise, no effect.\end{tabular}          \\ \hline
LATERAL MOVEMENT & \textit{\begin{tabular}[c]{@{}l@{}}non-compromised node \\ adjacent to stolen credentials\end{tabular}} & \begin{tabular}[c]{@{}l@{}}The attacker moves to a connected node using stolen credentials.\end{tabular} & The attacker gains access to the node. \\ \hline
\end{tabular}%
}
\end{table*}




%
The attacker is characterized by a list of products they are capable of targeting and a list of CVEs they are able to exploit. This reflects real APT behavior, as such groups are typically specialized in attacking products by specific vendors, and leverage publicly-available exploits, including those supported by tools such as Metasploit \cite{metasploit} and Nuclei \cite{projectdiscovery_nuclei}, and against vulnerabilities in the KEV catalog \cite{CISAKEV}.


The attacker can be configured to start from outside or inside the network, modeling exploitation of Internet-exposed vulnerable services and phishing-derived initial access, respectively. These correspond to the two most common vectors for achieving Initial Access \cite{mitre_attack_ta0001} in enterprise networks \cite{ArcticWolf2025_ThreatReport, Coalition2025_CyberThreatIndex, Sophos2025_ActiveAdversaryReport}.
If outside the network, the initial node is selected from those exposed to the Internet. If the attacker has phishing capabilities, the initial node of the attack is selected from within the network. 
%
Once the attacker is inside the network, it attempts to move laterally across neighboring nodes.

The set of actions available to the attacker depends on whether a node has already been compromised. 
On non-compromised nodes, the attacker may perform one of the following actions: \textit{Scan}, \textit{Exploit}, or \textit{DoS}. 
If the exploit is successful, the attacker gains access to the node, which becomes compromised. On compromised nodes, the attacker can execute post-exploitation actions, which include \textit{Credential Theft}, \textit{Privilege Escalation}, \textit{Persistence}, \textit{Data Exfiltration}, or \textit{Wiper}, each representing a distinct tactic aligned with common adversarial behaviors in real-world attacks.
If \textit{Credential Theft} is successful on a compromised node, the attacker extracts authentication material (e.g., credentials or tokens) that can be reused to perform the \textit{Lateral Movement} action. This enables the attacker to access other nodes in the network that are reachable from the compromised node, even if they are not directly vulnerable to exploitation.
If the defender applies a patch to a compromised node, the attacker is evicted from that node, unless a persistence action has been performed on the node.

Actions have a probabilistic outcome, reflecting real-world uncertainty. The probability of success can be configured based on Cyber Threat Intelligence (CTI) describing the attacker’s capabilities; for instance, the likelihood of successfully exploiting a vulnerability may vary depending on whether the attacker is specialized on the vulnerable product, whether the vulnerability is known to be exploited in the wild, and other characteristics (e.g., complexity of the vulnerability).

The selection of the target node, the action to perform, and the vulnerability to exploit is not predefined, but learned over time from simulation episodes. The attacker progressively constructs a policy that guides actions to maximize the impact of the attack campaign. 
The learning process is driven by a \emph{reward} structure, which assigns numerical values to the outcomes of different actions. The reward is defined as:
\begin{equation}
    R = A \times I
\end{equation} 
\noindent 
More specifically, $A$ is the importance of the action, which reflects the relevance of a specific action with respect to the attacker’s objective. For instance, actions that directly achieve the attacker’s goals (e.g., impact operations such as \textit{Data Exfiltration} and \textit{Wiper}) are assigned higher importance values, while information-gathering actions (e.g., \textit{Scan} and \textit{Credential Theft}) are associated with lower values; other actions fall between these extremes. 
$I$ denotes the importance of the node, defined in terms of its proximity to sensitive assets, with higher values assigned to nodes closer to critical resources.

Rewards are positive for successful actions. Conversely, a fixed negative reward is given for failed actions, since they do not contribute to the attacker's objective and represent wasted effort. 
%
The agent adopts an $\epsilon$-greedy action selection strategy to balance exploration and exploitation. With probability $\epsilon$, the attacker explores the action space by selecting a random move; with probability $1 - \epsilon$, it exploits its current knowledge by choosing the action that maximizes the estimated \textit{Q-value}.
To manage the transition from exploration to exploitation, an $\epsilon$-decay mechanism is employed. 

\subsection{Defenders}

The objective of the \textit{defender} agent is to protect the network by patching vulnerabilities, by following a vulnerability management policy under evaluation. Due to resource limitations, the defender must allocate efforts strategically.

The defender performs periodic vulnerability management cycles, as typical of enterprise systems \cite{purplesec_vulnerability,wiz_vulnerability,sentinelone_vulnerability}. A cycle consists of two phases: (i) the \textit{discovery phase} (which takes a fixed amount of time), in which the defender performs a scan to identify new vulnerabilities, reviews the results, and updates a priority queue of vulnerabilities (which includes both previous vulnerabilities that are still pending, and the new ones); and (ii) the \textit{resolution phase} (which takes a variable amount of time, as discussed in this section), in which the defender performs patching, by following the prioritization established by the discovery phase.



The cost required to patch a vulnerability represents the amount of time to apply a patch, in discrete time steps of the simulation (e.g., days). 
We model this cost as a function of multiple factors:

\begin{equation}
\label{eq:time_to_fix}
    T_{\text{fix}} = \alpha \cdot \beta \cdot \gamma \cdot \delta
\end{equation}
\noindent  
where:

\begin{itemize}
    \setlength{\itemsep}{0.6em}
    \setlength{\topsep}{0.6em}
    \setlength{\parsep}{0em}

    \item $\alpha$ represents the baseline patching time, i.e., the average time to patch a vulnerability in the simplest case, on a single node. This parameter is calibrated by the user based on the vulnerability management capabilities of the organization.

    \item $\beta = \frac{1}{\textit{defender effort}}$, where \textit{defender effort} is a scale parameter that captures the amount of resources allocated to vulnerability management, e.g., in terms of man-days. Higher effort corresponds to lower values of $\beta$, resulting in faster patching times. In this work, we consider three levels for the defender effort (\textit{Low} $< 1$, \textit{Regular} $= 1$, \textit{High} $> 1$).

    \item $\gamma$ is a scale parameter that represents an additional patching effort due to the type of vulnerability. In this work, we adopt two levels: $\gamma_{\text{APP}}$ for application-level vulnerabilities ($\gamma = 1$) and $\gamma_{\text{OS}}$ ($\gamma > 1$) for operating-system vulnerabilities. The latter are assumed to require longer patching times due to higher complexity.

    \item $\delta = \kappa \cdot |\mathcal{N}_v|$ is a scale parameter that represents an additional patching effort due to the number of affected nodes.  $|\mathcal{N}_v|$ is the number of nodes affected by the vulnerability. 
    $\kappa$ is a weight that reflects the impact of the number of nodes on $T_{\text{fix}}$. For example, it can be calibrated such that $T_{\text{fix}}$ matches the worst-case expected cost for deploying a patch over the entire network ($\mathcal{N}_v = \mathcal{N}$).
\end{itemize}
 
The cost to patch has been designed to be easy to interpret and configure by the user of \toolname{}. Since the time to apply a patch depends on the specific organization, we designed the factors above such that the user can apply knowledge from past vulnerability management activities, such as the average time to patch a simple one-node vulnerability, and to patch a vulnerability that spans the entire network. The formula can be easily modified to reflect different criteria, such as increasing $\delta$ with a logarithmic trend instead of a linear one.

\subsection{Implementation Details}
\toolname{} is implemented in Python and builds on \texttt{NetworkX}~\cite{networkx} for network modeling and \texttt{Gymnasium}~\cite{Gymnasium} for the RL environment. The framework is modular, with configuration files defining the network topology, vulnerability distribution, and agent settings. Experiments are run via Jupyter notebooks, which are used to produce all reported figures and tables.
\section{Experiments Design}
\label{sec:experimental_design}

In the rest of the paper, we present experiments based on \toolname{}, in order to show how the tool can be used for analyzing vulnerability management policies. We consider popular network topologies, and analyze real APT profiles according to threat intelligence sources. We remark that \toolname{} is a customizable tool, which can be configured by security analysts to simulate specific network environments and threats. Thus, experimental results should be interpreted in the light of the specific context of interest. We release the full source code of the tool and the configuration of the experiments, in order to support reproducibility and customizations.

In particular, \toolname{} produces the following metrics, which will be discussed in Section~\ref{sec:evaluation}:

\begin{itemize}
    \item \textit{Goal Achievement Rate}: the percentage of episodes in which the attacker successfully reaches the campaign goal;
    \item \textit{Network Vulnerability Index } (NVI): the percentage of network nodes to which the attacker has gained access, relative to the total number of nodes in the network;
    \item \textit{Time to Reach Goal} (TTRG): the number of simulated days required for the attacker to reach its goal;
    \item \textit{Time to Patch Vulnerability}(TTPV): the average number of days a policy takes to remediate a vulnerability since it is known;
    \item \textit{Vulnerabilities in Backlog} (VIB): the average number of vulnerabilities awaiting remediation at any given time.
\end{itemize}

%
%

\begin{figure}[htbp]
    \centering
    \includegraphics[width=0.7\columnwidth]{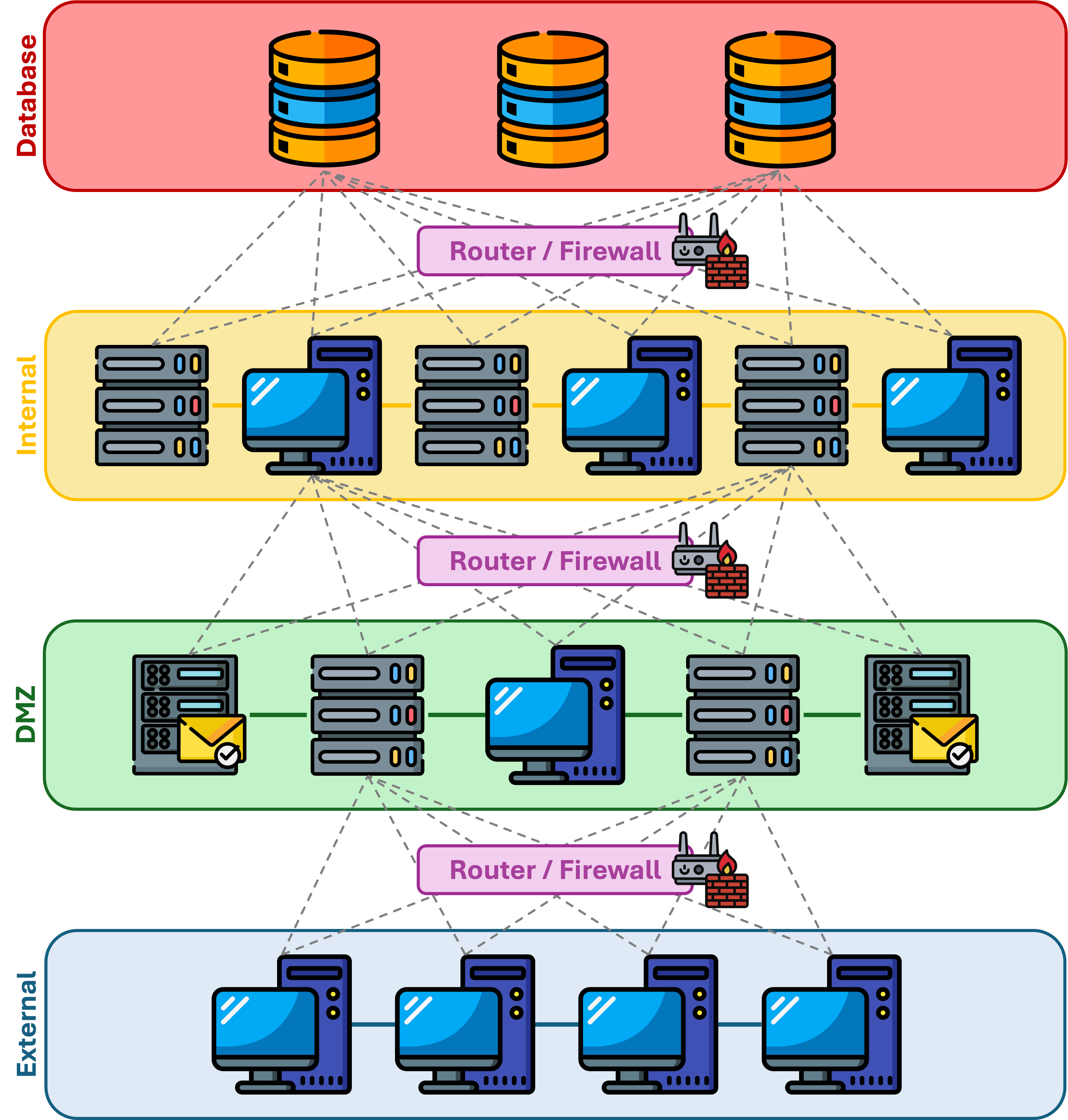}
    \caption{Layered network topology.}
    \label{fig:layered_topology}
\end{figure}

\subsection{Simulated Network}

We conducted the experiments on two network topologies of equal size, each comprising $100$ nodes, differing only in their structural organization, in order to enable a controlled comparison between topologies. This network size is consistent with typical Small and Medium Enterprise (SME) deployments, which are generally reported to range from a few dozen to a few hundred networked devices~\cite{10.1145/3776942.3777008, 10.1145/3773365.3773603}.
The network is organized into multiple security layers (i.e., External, DMZ, Internal, and Database), which are interconnected through routing and firewall devices. These components regulate and filter traffic between layers. As a result, only a subset of nodes in each layer is reachable from other layers, reflecting realistic network segmentation and restricted connectivity patterns.

\vspace{3pt}
\noindent
\textbf{\textit{Layered} topology} (Figure~\ref{fig:layered_topology}). In this configuration, nodes are organized into multiple layers, where only a subset of nodes in each layer is reachable from the previous one. These nodes act as shared entry points between layers, such as web servers or other exposed services.
This topology reflects real-world scenarios such as enterprise networks composed of multiple subnets (e.g., DMZ, employee network, and shared resources network) \cite{oracle_dmz}, as well as industrial environments with segmented architectures (e.g., DMZ, IT, and OT networks) \cite{industrial_cybersec}.

\begin{figure}[htbp]
    \centering
    \includegraphics[width=0.7\columnwidth]{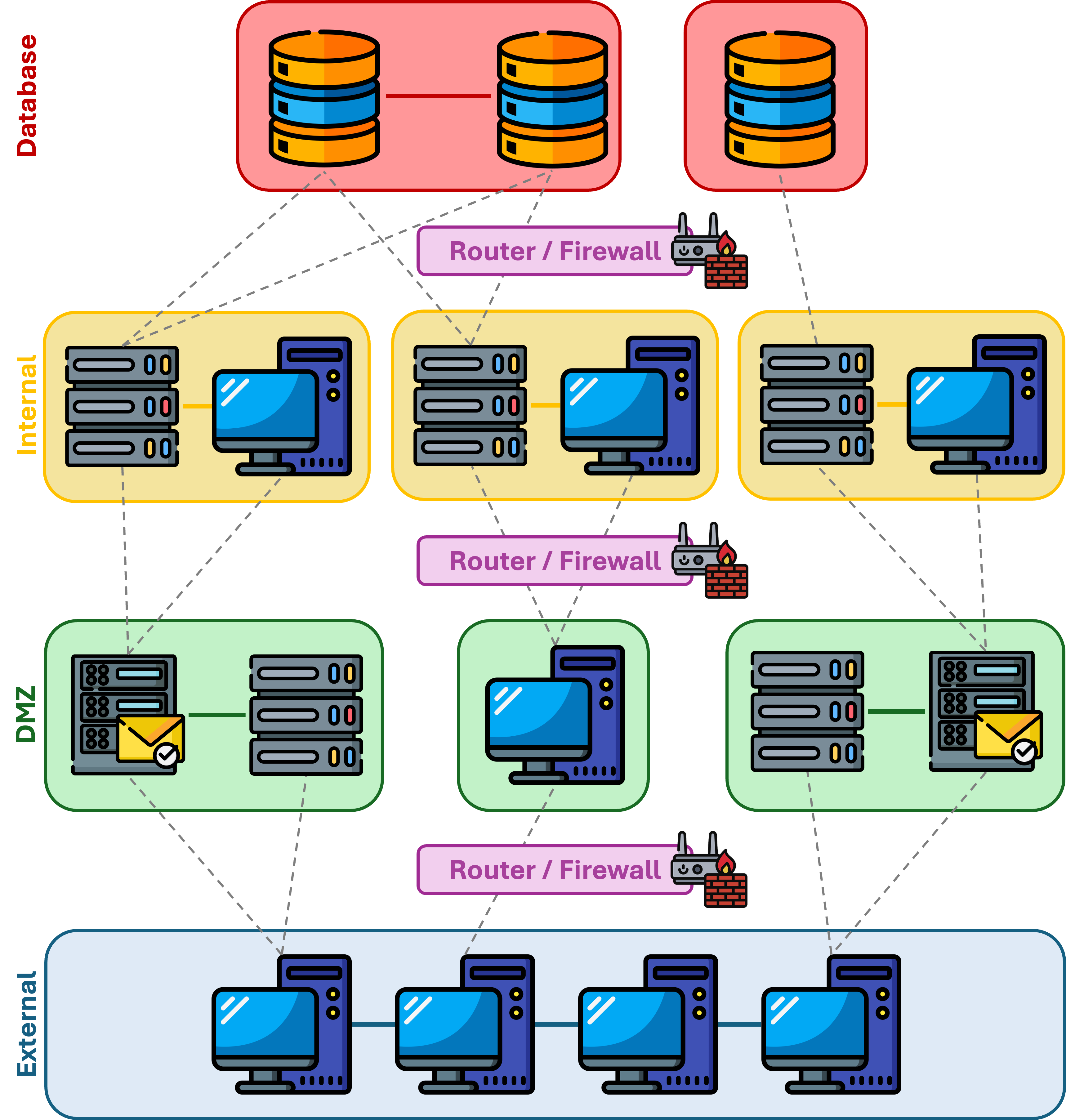}
    \caption{Tree network topology.}
    \label{fig:tree_topology}
\end{figure}

\noindent
\textbf{\textit{Tree} topology} (Figure~\ref{fig:tree_topology}). In this configuration, the network is structured hierarchically, with a primary network connected to multiple secondary networks, and optionally to additional lower-level networks. These networks are interconnected through a subset of nodes belonging to the higher-level network.
This structure models real-world scenarios with geographically-distributed enterprise networks, such as in the IBM Hierarchical Tree Topology \cite{ibm_topology}.



The simulation models an enterprise environment in which each zone is associated with a set of representative software products. For each layer, products are sampled from predefined product categories to ensure diversity while maintaining a realistic distribution. The External zone includes commonly used client-side applications and operating systems (e.g., Adobe products, Microsoft browsers and tools, and Windows). The DMZ comprises network-facing services, including web and application servers, remote access components, virtualization platforms, and Linux-based operating systems. The Internal zone encompasses a broad range of enterprise technologies, including operating systems (e.g., Windows, Debian, and Red Hat), and infrastructure and virtualization services (e.g., VMware). Finally, the Database zone consists of database management systems, including relational solutions (e.g., MariaDB, PostgreSQL, Oracle, and SQLite) as well as NoSQL systems (e.g., MongoDB).
Vulnerabilities affecting these products are collected from the NVD~\cite{nist_nvd}. We considered the ones disclosed in the year $2020$, resulting in $4{,}137$ vulnerabilities. To ensure scalability and computational efficiency, stratified sampling is applied to the set of vulnerabilities, preserving their distribution across products and severity levels. The resulting subset is mapped onto a fixed network of 100 nodes, where each node may host multiple vulnerabilities, with a maximum of 10 vulnerabilities per node. 
Simulation episodes represent a full one-year time window. 

\subsection{Agents}\label{sec:agents}

\subsubsection{Attacker}

In the following experiments, we evaluate the impact of APT campaigns on the target network. 
We consider two common types of campaign: \textit{Exfiltration/Wiper} and \textit{Denial of Service (DoS)}. In the Exfiltration/Wiper campaign, the goal is to perform data exfiltration and destructive actions on database nodes; in the DoS campaign, the objective is to disrupt availability of database nodes. We use \toolname{} to measure the likelihood of success of the attack campaigns over multiple runs of the simulation. In these experiments, we define campaign success as Exfiltration/Wiper or DoS on at least three database nodes, to represent multi-node impact on critical assets. The threshold is configurable.

%

The attacker's behavior is modeled after two distinct APT profiles: APT41~\cite{mitre-apt41} for the Exfiltration/Wiper campaign, and APT28~\cite{mitre-apt28} for the DoS campaign. 
In each campaign, the attacker is assumed to be specialized in a set of products, and can therefore exploit only vulnerabilities affecting those products, as indicated by CTI sources. We derive this information from the CTI‑HAL dataset \cite{cti-hal}. We configure the attacker to also exploit vulnerabilities listed in the KEV catalog~\cite{CISAKEV}. 
Furthermore, both APT41 and APT28 possess phishing capabilities; consequently, Initial Access can occur on both external and internal nodes.


Hyperparameters for RL were configured as  in Table \ref{tab:dqn_atc_parameters}, based on prior research applying RL to cybersecurity domains \cite{wiebe2023learning}. 
The attacker agent was trained in a standalone setting (i.e., without a defender) through a dedicated training phase of $1{,}000$ simulation episodes. Training was performed separately for each combination of APT and network topology, resulting in independent training runs for all APT–topology pairs.

\begin{table}[htbp]
\caption{RL hyperparameters.}
\begin{tabular}{|l|l|l|}
\hline
\textbf{Parameter} & \textbf{Description} & \textbf{Set Value} \\ \hline
\texttt{alpha} & Learning rate. & $0.001$ \\ \hline
\texttt{gamma} & Discount factor. & $0.99$ \\ \hline
\texttt{epsilon} & Exploration probability. & $1.0$ \\ \hline
\texttt{epsilon\_decay} & Epsilon decay rate. & $0.995$ \\ \hline
\texttt{epsilon\_min} & Minimum value for epsilon. & $0.01$ \\ \hline
\texttt{memory\_size} & Replay buffer size. & $10,000$ \\ \hline
{\texttt{batch\_size}} & Number of experiences sampled to update the neural network. & {$64$} \\ \hline
\end{tabular}
\label{tab:dqn_atc_parameters}
\end{table}

\subsubsection{Defender}


%
The defender alternates between discovery and remediation activities. Periodically every week, the defender performs discovery, in which new vulnerabilities are identified. The following days of the cycle are dedicated to remediation, in which the defender sequentially patches vulnerabilities, according to the priority defined by the policy under evaluation. If a higher-priority vulnerability arises while another one is being patched, the remediation activity is not preempted in order to avoid wasting efforts. The tool also allows the user to enable preemption.

We considered three vulnerability prioritization policies. \textit{Policy-Severity} prioritizes vulnerabilities based on their CVSS severity score, as reported in the NVD \cite{CVSS}. \textit{Policy-Importance} ranks vulnerabilities according to the average importance of the nodes they affect, prioritizing those impacting more critical assets. \textit{Policy-Centrality} prioritizes vulnerabilities affecting nodes with higher network centrality, capturing structural importance within the system graph and reflecting alternative interpretations of asset criticality \cite{farris2018vulcon}.

We configured the parameters of the time-to-patch as follows (Eq.~\ref{eq:time_to_fix}). The baseline patching time $\alpha$ is set to $1$ day. The defender efficiency factor $\beta$ is set to $0.5$ for high effort, $1$ for regular effort and $1.5$ for low effort. The parameter $\gamma$ is set to $1$ for application-level vulnerabilities, and to $2$ for OS ones, reflecting the higher complexity and operational impact of patching OS components. The weight $\kappa$ that defines $\delta$ is set to $0.3$ with linear growth, accounting for parallelization effects and operational efficiencies. Overall, this parameterization results in patching times ranging approximately from $1$ day up to about $12$ days in the worst case.



\section{Evaluation}
\label{sec:evaluation}

We analyze multiple scenarios, by considering every combination of the two APTs and of the two topologies. For each case, we analyze the three patching policies (\emph{Severity}, \emph{Importance}, \emph{Centrality}), with the three configurations of the defender effort, and the case of no defender (\emph{None}).
Each scenario was executed over $100$ independent episodes to ensure statistical significance. Experiments were run on a MacOS machine with M2 Pro processor and $16$ GB RAM. On this hardware, agent training took approximately $4$ minutes, while the full execution took approximately $12$ minutes.

Table~\ref{tab:metrics_apt41_apt28} reports the percentage of episodes in which the attacker successfully reaches its goal (respectively, Exfiltration/Wiper and DoS on at least 3 nodes in the Database layer). 
In the absence of a defender, both attackers consistently achieve their objectives in all settings.
When defenses are introduced, \textit{Policy-Importance} provides the strongest mitigation effect. This is particularly evident for APT41 in the Layered topology, where increasing defender effort leads to a near-complete suppression of attacker success. For APT28, the reduction is less pronounced, with the attacker remaining more effective even under higher effort levels.
This difference can be partly explained by the nature of the two campaigns, since APT28 pursues a simpler objective that requires only a single action (DoS) on the database nodes, while APT41 entails multiple actions to impact on these nodes, making it more sensitive to defensive interventions.
In contrast, \textit{Severity} and \textit{Centrality} policies show limited impact across both campaigns, even at higher effort levels, with only marginal reductions in attacker performance. This is likely due to their focus on criteria not directly aligned with protecting the assets targeted by the attacker. Conversely, \textit{Importance} explicitly prioritizes those assets that are most relevant to the attacker’s objectives, which explains its higher effectiveness.

\begin{table}[htbp]
\centering
\caption{Attacker Goal Achievement (\%).}
\label{tab:metrics_apt41_apt28}
\begin{tabular}{|l|l|rrrr|}
\hline
\multicolumn{1}{|c|}{\multirow{4}{*}{\textbf{Defender}}} &
  \multicolumn{1}{c|}{\multirow{4}{*}{\textbf{Effort}}} &
  \multicolumn{4}{c|}{\textbf{APT}} \\ \cline{3-6} 
\multicolumn{1}{|c|}{} &
  \multicolumn{1}{c|}{} &
  \multicolumn{2}{c|}{\textit{APT41}} &
  \multicolumn{2}{c|}{\textit{APT28}} \\ \cline{3-6} 
\multicolumn{1}{|c|}{} &
  \multicolumn{1}{c|}{} &
  \multicolumn{4}{c|}{\textbf{Topology}} \\ \cline{3-6} 
\multicolumn{1}{|c|}{} &
  \multicolumn{1}{c|}{} &
  \multicolumn{1}{c|}{\textit{Layered}} &
  \multicolumn{1}{c|}{\textit{Tree}} &
  \multicolumn{1}{c|}{\textit{Layered}} &
  \multicolumn{1}{c|}{\textit{Tree}} \\ \hline
None &
  - &
  \multicolumn{1}{r|}{100 \DrawPercentageBar{1.00}} &
  \multicolumn{1}{r|}{100 \DrawPercentageBar{1.00}} &
  \multicolumn{1}{r|}{100 \DrawPercentageBar{1.00}} &
  100 \DrawPercentageBar{1.00} \\ \hline
\multirow{3}{*}{Importance} &
  \textit{Low} &
  \multicolumn{1}{r|}{56 \DrawPercentageBar{0.56}} &
  \multicolumn{1}{r|}{48 \DrawPercentageBar{0.48}} &
  \multicolumn{1}{r|}{98 \DrawPercentageBar{0.98}} &
  98 \DrawPercentageBar{0.98} \\ \cline{2-6} 
 &
  \textit{Regular} &
  \multicolumn{1}{r|}{30 \DrawPercentageBar{0.30}} &
  \multicolumn{1}{r|}{16 \DrawPercentageBar{0.16}} &
  \multicolumn{1}{r|}{94 \DrawPercentageBar{0.94}} &
  93 \DrawPercentageBar{0.93} \\ \cline{2-6} 
 &
  \textit{High} &
  \multicolumn{1}{r|}{3 \DrawPercentageBar{0.03}} &
  \multicolumn{1}{r|}{0 \DrawPercentageBar{0.00}} &
  \multicolumn{1}{r|}{81 \DrawPercentageBar{0.81}} &
  57 \DrawPercentageBar{0.57} \\ \hline
\multirow{3}{*}{Severity} &
  \textit{Low} &
  \multicolumn{1}{r|}{100 \DrawPercentageBar{1.00}} &
  \multicolumn{1}{r|}{100 \DrawPercentageBar{1.00}} &
  \multicolumn{1}{r|}{99 \DrawPercentageBar{0.99}} &
  98 \DrawPercentageBar{0.98} \\ \cline{2-6} 
 &
  \textit{Regular} &
  \multicolumn{1}{r|}{100 \DrawPercentageBar{1.00}} &
  \multicolumn{1}{r|}{100 \DrawPercentageBar{1.00}} &
  \multicolumn{1}{r|}{99 \DrawPercentageBar{0.99}} &
  97 \DrawPercentageBar{0.97} \\ \cline{2-6} 
 &
  \textit{High} &
  \multicolumn{1}{r|}{83 \DrawPercentageBar{0.83}} &
  \multicolumn{1}{r|}{72 \DrawPercentageBar{0.72}} &
  \multicolumn{1}{r|}{94 \DrawPercentageBar{0.94}} &
  90 \DrawPercentageBar{0.90} \\ \hline
\multirow{3}{*}{Centrality} &
  \textit{Low} &
  \multicolumn{1}{r|}{100 \DrawPercentageBar{1.00}} &
  \multicolumn{1}{r|}{100 \DrawPercentageBar{1.00}} &
  \multicolumn{1}{r|}{100 \DrawPercentageBar{1.00}} &
  96 \DrawPercentageBar{0.96} \\ \cline{2-6} 
 &
  \textit{Regular} &
  \multicolumn{1}{r|}{100 \DrawPercentageBar{1.00}} &
  \multicolumn{1}{r|}{100 \DrawPercentageBar{1.00}} &
  \multicolumn{1}{r|}{100 \DrawPercentageBar{1.00}} &
  96 \DrawPercentageBar{0.96} \\ \cline{2-6} 
 &
  \textit{High} &
  \multicolumn{1}{r|}{92 \DrawPercentageBar{0.92}} &
  \multicolumn{1}{r|}{81 \DrawPercentageBar{0.81}} &
  \multicolumn{1}{r|}{99 \DrawPercentageBar{0.99}} &
  95 \DrawPercentageBar{0.95} \\ \hline
\end{tabular}
\end{table}

\begin{figure*}[ht]
     \centering
     \subfloat[{Attacker vs. Policy-Importance}\label{fig:heatmap_atk}]{\includegraphics[height=.28\textwidth]{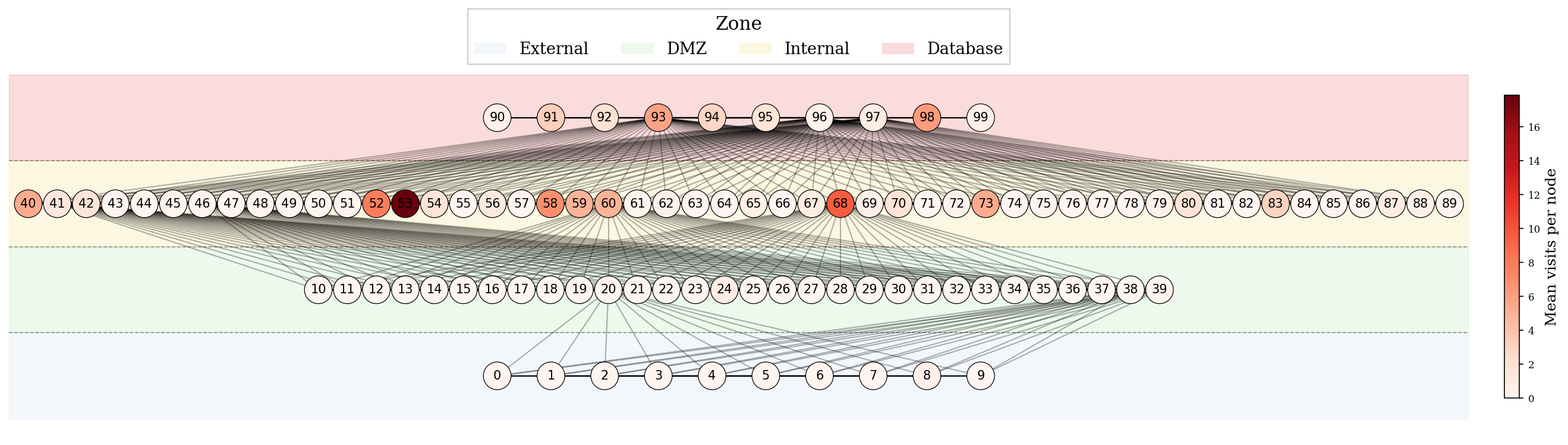}}
     \hfill
     \subfloat[{Defender with \textit{Policy-Importance}}\label{fig:heatmap_def_importance}]
     {\includegraphics[height=.28\textwidth]{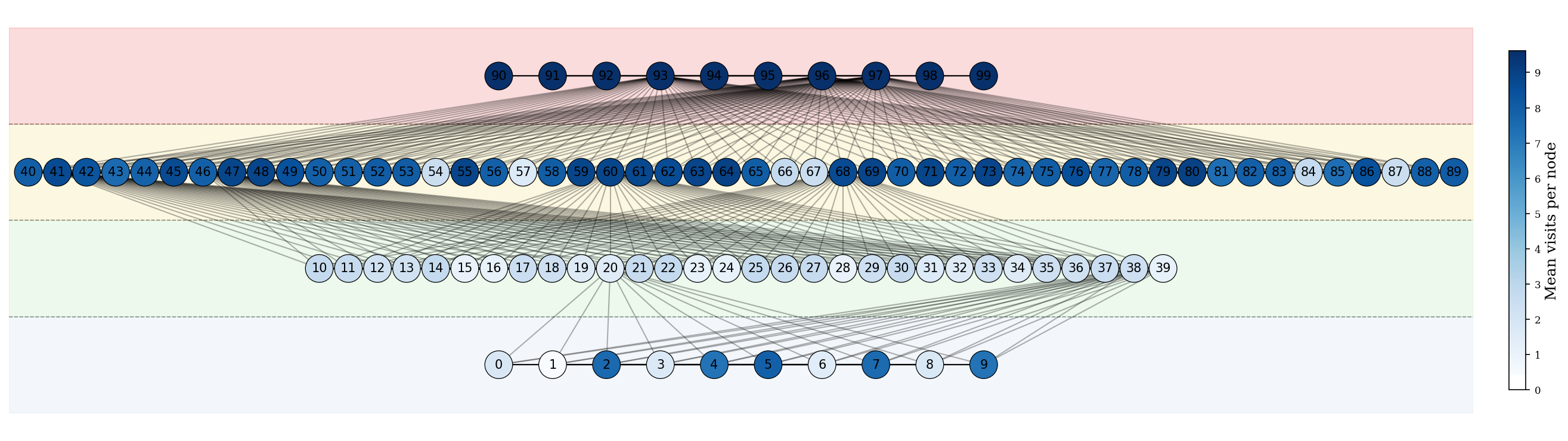}}
          \hfill
     \subfloat[{Defender with \textit{Policy-Severity}}\label{fig:heatmap_def_severity}]
     {\includegraphics[height=.28\textwidth]{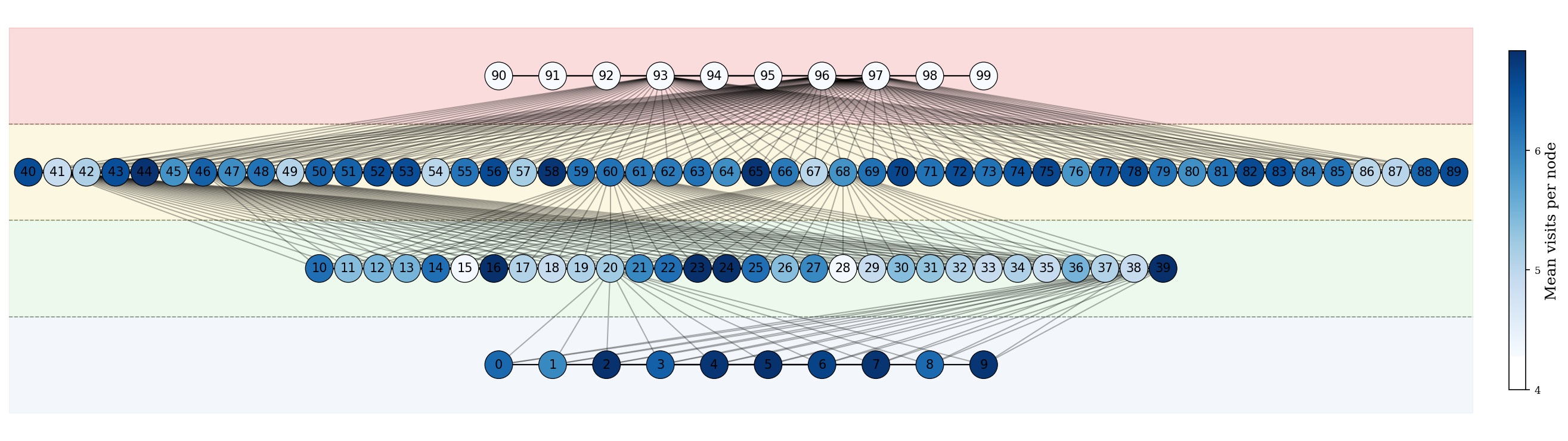}}
    \caption{Heatmaps of attacker and defender node targets in a layered network topology.}
    \label{fig:heatmaps}
\end{figure*}

To provide a deeper qualitative view of attacker and defender behavior, \toolname{} produces a heatmap of the network in which the color intensity of each node reflects the average number of times that node was selected as the target of an action across episodes (Figure~\ref{fig:heatmaps}).
As shown in Figure~\ref{fig:heatmap_atk}, the attacker tends to traverse specific intermediate nodes to progress toward the final objective. For example, node $68$ emerges as a key pivot point. This suggests that certain nodes are systematically exploited as stepping stones during the attack progression. Furthermore, several nodes within the database zone exhibit higher intensities, confirming that the attacker successfully reaches its objective.
Figure~\ref{fig:heatmap_def_importance} illustrates the behavior of the defender under \textit{Policy-Importance} strategy. The defender indeed prioritizes critical assets, by focusing patching efforts on the database zone. The color intensity gradually fades as the distance from the database zone increases, showing that nodes distant from critical assets are patched less frequently.
In contrast, Figure~\ref{fig:heatmap_def_severity} depicts the defender’s behavior under \textit{Policy-Severity} strategy, which targets vulnerabilities with the highest CVSS scores. The comparison highlights that addressing the most ``severe'' vulnerabilities does not necessarily protect critical resources. In this scenario, the database zone and pivotal nodes (e.g., node $68$) receive less emphasis from the defender, which ultimately facilitates the attacker in reaching its goal. This limited effectiveness can be attributed to the fact that severity-based prioritization does not fully capture the characteristics of real-world exploited vulnerabilities, as reflected in the KEV catalog, where factors beyond CVSS scores play a crucial role. 
Moreover, we note that the policy gives higher priority to nodes in the external layer, which is less effective at protecting from attacks that originate from within the network due to phishing.


\toolname{} also reports several other metrics to support a deeper analysis of policies.

The \textit{Network Vulnerability Index} (NVI) represents the percentage of network nodes to which the attacker has gained access, relative to the total number of nodes in the network. Table~\ref{tab:metrics_nvi} reports the corresponding values for each configuration.
The results show that NVI generally decreases as the defender's \textit{effort} increases, reflecting a reduction in the overall fraction of compromised nodes. However, minimizing the number of compromised nodes does not necessarily make the attack more difficult, as patching may protect nodes that are of limited relevance to the attacker's objective. This indicates that NVI captures the overall exposure of the network, rather than the attacker's progress toward its specific target. Accordingly, policy effectiveness should be assessed with respect to the goal achievement rate, rather than the raw number of protected nodes.

\begin{table}[htbp]
\centering
\caption{Mean Network Vulnerability Index (\%)}
\label{tab:metrics_nvi}
\begin{tabular}{|l|l|rrrr|}
\hline
\multicolumn{1}{|c|}{\multirow{4}{*}{\textbf{Defender}}} &
  \multicolumn{1}{c|}{\multirow{4}{*}{\textbf{Effort}}} &
  \multicolumn{4}{c|}{\textbf{APT}} \\ \cline{3-6} 
\multicolumn{1}{|c|}{} &
  \multicolumn{1}{c|}{} &
  \multicolumn{2}{c|}{\textit{APT41}} &
  \multicolumn{2}{c|}{\textit{APT28}} \\ \cline{3-6} 
\multicolumn{1}{|c|}{} &
  \multicolumn{1}{c|}{} &
  \multicolumn{4}{c|}{\textbf{Topology}} \\ \cline{3-6} 
\multicolumn{1}{|c|}{} &
  \multicolumn{1}{c|}{} &
  \multicolumn{1}{c|}{\textit{Layered}} &
  \multicolumn{1}{c|}{\textit{Tree}} &
  \multicolumn{1}{c|}{\textit{Layered}} &
  \multicolumn{1}{c|}{\textit{Tree}} \\ \hline
None &
  - &
  \multicolumn{1}{r|}{34 \DrawPercentageBar{0.34}} &
  \multicolumn{1}{r|}{39 \DrawPercentageBar{0.39}} &
  \multicolumn{1}{r|}{29 \DrawPercentageBar{0.29}} &
  22 \DrawPercentageBar{0.22} \\ \hline
\multirow{3}{*}{Importance} &
  \textit{Low} &
  \multicolumn{1}{r|}{21 \DrawPercentageBar{0.21}} &
  \multicolumn{1}{r|}{19 \DrawPercentageBar{0.19}} &
  \multicolumn{1}{r|}{18 \DrawPercentageBar{0.18}} &
  17 \DrawPercentageBar{0.17} \\ \cline{2-6} 
 &
  \textit{Regular} &
  \multicolumn{1}{r|}{17 \DrawPercentageBar{0.17}} &
  \multicolumn{1}{r|}{14 \DrawPercentageBar{0.14}} &
  \multicolumn{1}{r|}{13 \DrawPercentageBar{0.13}} &
  14 \DrawPercentageBar{0.14} \\ \cline{2-6} 
 &
  \textit{High} &
  \multicolumn{1}{r|}{19 \DrawPercentageBar{0.19}} &
  \multicolumn{1}{r|}{12 \DrawPercentageBar{0.12}} &
  \multicolumn{1}{r|}{6 \DrawPercentageBar{0.06}} &
  8 \DrawPercentageBar{0.08} \\ \hline
\multirow{3}{*}{Severity} &
  \textit{Low} &
  \multicolumn{1}{r|}{23 \DrawPercentageBar{0.23}} &
  \multicolumn{1}{r|}{26 \DrawPercentageBar{0.26}} &
  \multicolumn{1}{r|}{19 \DrawPercentageBar{0.19}} &
  21 \DrawPercentageBar{0.21} \\ \cline{2-6} 
 &
  \textit{Regular} &
  \multicolumn{1}{r|}{22 \DrawPercentageBar{0.22}} &
  \multicolumn{1}{r|}{25 \DrawPercentageBar{0.25}} &
  \multicolumn{1}{r|}{19 \DrawPercentageBar{0.19}} &
  19 \DrawPercentageBar{0.19} \\ \cline{2-6} 
 &
  \textit{High} &
  \multicolumn{1}{r|}{15 \DrawPercentageBar{0.15}} &
  \multicolumn{1}{r|}{13 \DrawPercentageBar{0.13}} &
  \multicolumn{1}{r|}{7 \DrawPercentageBar{0.07}} &
  9 \DrawPercentageBar{0.09} \\ \hline
\multirow{3}{*}{Centrality} &
  \textit{Low} &
  \multicolumn{1}{r|}{25 \DrawPercentageBar{0.25}} &
  \multicolumn{1}{r|}{29 \DrawPercentageBar{0.29}} &
  \multicolumn{1}{r|}{18 \DrawPercentageBar{0.18}} &
  19 \DrawPercentageBar{0.19} \\ \cline{2-6} 
 &
  \textit{Regular} &
  \multicolumn{1}{r|}{23 \DrawPercentageBar{0.23}} &
  \multicolumn{1}{r|}{25 \DrawPercentageBar{0.25}} &
  \multicolumn{1}{r|}{16 \DrawPercentageBar{0.16}} &
  18 \DrawPercentageBar{0.18} \\ \cline{2-6} 
 &
  \textit{High} &
  \multicolumn{1}{r|}{16 \DrawPercentageBar{0.16}} &
  \multicolumn{1}{r|}{12 \DrawPercentageBar{0.12}} &
  \multicolumn{1}{r|}{8 \DrawPercentageBar{0.08}} &
  8 \DrawPercentageBar{0.08} \\ \hline
\end{tabular}
\end{table}

The \textit{Time to Reach Goal} (TTRG) represents the number of simulated days required for the attacker to reach its goal. Table~\ref{tab:mean_ttrg} reports the corresponding values for each configuration.
Higher TTRG values indicate that the attacker faces greater difficulty in succeeding, thereby providing defenders with a wider time window to detect the ongoing attack.
The results show that, as the defender's \textit{effort} increases, the system's defensive capability improves, requiring the attacker a greater number of days to reach the goal. In particular, \textit{Policy-Importance} yields a significant increase in TTRG compared to the other policies.
The difference in the number of days between the two campaigns is due to their different nature, as already discussed previously.

\begin{table}[htbp]
\centering
\caption{Mean Time To Reach the Goal (days)}
\label{tab:mean_ttrg}
\begin{tabular}{|l|l|cccc|}
\hline
\multicolumn{1}{|c|}{\multirow{4}{*}{\textbf{Defender}}} &
  \multicolumn{1}{c|}{\multirow{4}{*}{\textbf{Effort}}} &
  \multicolumn{4}{c|}{\textbf{APT}} \\ \cline{3-6} 
\multicolumn{1}{|c|}{}      & \multicolumn{1}{c|}{} & \multicolumn{2}{c|}{\textit{APT41}}                 & \multicolumn{2}{c|}{\textit{APT28}} \\ \cline{3-6} 
\multicolumn{1}{|c|}{}      & \multicolumn{1}{c|}{} & \multicolumn{4}{c|}{\textbf{Topology}}                                                    \\ \cline{3-6} 
\multicolumn{1}{|c|}{} &
  \multicolumn{1}{c|}{} &
  \multicolumn{1}{c|}{\textit{Layered}} &
  \multicolumn{1}{c|}{\textit{Tree}} &
  \multicolumn{1}{c|}{\textit{Layered}} &
  \textit{Tree} \\ \hline
None                        & -                     & \multicolumn{1}{c|}{125} & \multicolumn{1}{c|}{128} & \multicolumn{1}{c|}{104}    & 108   \\ \hline
\multirow{3}{*}{Importance} & \textit{Low}          & \multicolumn{1}{c|}{215} & \multicolumn{1}{c|}{237} & \multicolumn{1}{c|}{152}    & 163   \\ \cline{2-6} 
                            & \textit{Regular}      & \multicolumn{1}{c|}{243} & \multicolumn{1}{c|}{244} & \multicolumn{1}{c|}{171}    & 178   \\ \cline{2-6} 
                            & \textit{High}         & \multicolumn{1}{c|}{315} & \multicolumn{1}{c|}{270} & \multicolumn{1}{c|}{190}    & 202   \\ \hline
\multirow{3}{*}{Severity}   & \textit{Low}          & \multicolumn{1}{c|}{165} & \multicolumn{1}{c|}{180} & \multicolumn{1}{c|}{120}    & 133   \\ \cline{2-6} 
                            & \textit{Regular}      & \multicolumn{1}{c|}{178} & \multicolumn{1}{c|}{204} & \multicolumn{1}{c|}{135}    & 154   \\ \cline{2-6} 
                            & \textit{High}         & \multicolumn{1}{c|}{236} & \multicolumn{1}{c|}{242} & \multicolumn{1}{c|}{158}    & 187   \\ \hline
\multirow{3}{*}{Centrality} & \textit{Low}          & \multicolumn{1}{c|}{150} & \multicolumn{1}{c|}{160} & \multicolumn{1}{c|}{118}    & 132   \\ \cline{2-6} 
                            & \textit{Regular}      & \multicolumn{1}{c|}{158} & \multicolumn{1}{c|}{175} & \multicolumn{1}{c|}{125}    & 144   \\ \cline{2-6} 
                            & \textit{High}         & \multicolumn{1}{c|}{216} & \multicolumn{1}{c|}{223} & \multicolumn{1}{c|}{138}    & 173   \\ \hline
\end{tabular}
\end{table}

The \textit{Time to Patch Vulnerability} (TTPV) measures the average number of days that a policy takes to remediate a vulnerability since it is known. Table~\ref{tab:meanTTPV} reports the corresponding values for each configuration.
The results show that, as the defender's \textit{effort} increases, fewer days are required to patch a vulnerability: with \textit{Low} effort, approximately $10$ days are needed on average, whereas with \textit{High} effort this decreases to approximately $4$ days.
These results can be used as Key Performance Indicators for guiding vulnerability management: if the observed patching time falls within this range, the process can be considered adequate; otherwise, it may indicate an issue to be addressed.

\begin{table}[htbp]
\centering
\caption{Mean Time To Patch Vulnerability (days)}
\label{tab:meanTTPV}
\begin{tabular}{|l|l|cccc|}
\hline
\multicolumn{1}{|c|}{\multirow{4}{*}{\textbf{Defender}}} &
  \multicolumn{1}{c|}{\multirow{4}{*}{\textbf{Effort}}} &
  \multicolumn{4}{c|}{\textbf{APT}} \\ \cline{3-6} 
\multicolumn{1}{|c|}{}      & \multicolumn{1}{c|}{} & \multicolumn{2}{c|}{\textit{APT41}}               & \multicolumn{2}{c|}{\textit{APT28}} \\ \cline{3-6} 
\multicolumn{1}{|c|}{}      & \multicolumn{1}{c|}{} & \multicolumn{4}{c|}{\textbf{Topology}}                                                  \\ \cline{3-6} 
\multicolumn{1}{|c|}{} &
  \multicolumn{1}{c|}{} &
  \multicolumn{1}{c|}{\textit{Layered}} &
  \multicolumn{1}{c|}{\textit{Tree}} &
  \multicolumn{1}{c|}{\textit{Layered}} &
  \textit{Tree} \\ \hline
\multirow{3}{*}{Importance} & \textit{Low}          & \multicolumn{1}{c|}{13} & \multicolumn{1}{c|}{13} & \multicolumn{1}{c|}{9}      & 8     \\ \cline{2-6} 
                            & \textit{Regular}      & \multicolumn{1}{c|}{10} & \multicolumn{1}{c|}{10} & \multicolumn{1}{c|}{6}      & 6     \\ \cline{2-6} 
                            & \textit{High}         & \multicolumn{1}{c|}{4}  & \multicolumn{1}{c|}{4}  & \multicolumn{1}{c|}{3}      & 3     \\ \hline
\multirow{3}{*}{Severity}   & \textit{Low}          & \multicolumn{1}{c|}{10} & \multicolumn{1}{c|}{10} & \multicolumn{1}{c|}{8}      & 7     \\ \cline{2-6} 
                            & \textit{Regular}      & \multicolumn{1}{c|}{7}  & \multicolumn{1}{c|}{7}  & \multicolumn{1}{c|}{5}      & 5     \\ \cline{2-6} 
                            & \textit{High}         & \multicolumn{1}{c|}{4}  & \multicolumn{1}{c|}{3}  & \multicolumn{1}{c|}{3}      & 3     \\ \hline
\multirow{3}{*}{Centrality} & \textit{Low}          & \multicolumn{1}{c|}{10} & \multicolumn{1}{c|}{10} & \multicolumn{1}{c|}{8}      & 7     \\ \cline{2-6} 
                            & \textit{Regular}      & \multicolumn{1}{c|}{7}  & \multicolumn{1}{c|}{7}  & \multicolumn{1}{c|}{5}      & 5     \\ \cline{2-6} 
                            & \textit{High}         & \multicolumn{1}{c|}{4}  & \multicolumn{1}{c|}{3}  & \multicolumn{1}{c|}{3}      & 3     \\ \hline
\end{tabular}
\end{table}

The \textit{Vulnerabilities in Backlog} (VIB) reports the average number of vulnerabilities awaiting remediation at any given time. Table~\ref{tab:meanVIB} reports the corresponding values for each configuration.
The results show that, as the defender's \textit{effort} increases, the number of vulnerabilities in the backlog decreases: with \textit{Low} effort, approximately $15$ vulnerabilities remain in the backlog on average, whereas with \textit{High} effort this decreases to approximately $5$.
Notably, the average VIB remains well above zero across all configurations, indicating that the defender does not have sufficient budget to promptly patch all known vulnerabilities. This observation is consistent with what is commonly reported in practice by security analysts, who remark the need for effective policies for vulnerability prioritization~\cite{ponemon2019vulnresponse, qualys2026backlog, first2026forecast}.
However, a low backlog length is not synonymous with defensive effectiveness: even when few vulnerabilities remain in the backlog, the attacker may still be able to reach the goal in a large fraction of cases. For instance, under \textit{Policy-Centrality} with the APT41 campaign in the \textit{layered} topology, the attacker achieves the goal in $92\%$ of cases despite an average backlog of only $4$ vulnerabilities.

\begin{table}[htbp]
\centering
\caption{Mean Vulnerability In Backlog (\#)}
\label{tab:meanVIB}
\begin{tabular}{|l|l|cccc|}
\hline
\multicolumn{1}{|c|}{\multirow{4}{*}{\textbf{Defender}}} &
  \multicolumn{1}{c|}{\multirow{4}{*}{\textbf{Effort}}} &
  \multicolumn{4}{c|}{\textbf{APT}} \\ \cline{3-6} 
\multicolumn{1}{|c|}{}      & \multicolumn{1}{c|}{} & \multicolumn{2}{c|}{\textit{APT41}}               & \multicolumn{2}{c|}{\textit{APT28}} \\ \cline{3-6} 
\multicolumn{1}{|c|}{}      & \multicolumn{1}{c|}{} & \multicolumn{4}{c|}{\textbf{Topology}}                                                  \\ \cline{3-6} 
\multicolumn{1}{|c|}{} &
  \multicolumn{1}{c|}{} &
  \multicolumn{1}{c|}{\textit{Layered}} &
  \multicolumn{1}{c|}{\textit{Tree}} &
  \multicolumn{1}{c|}{\textit{Layered}} &
  \textit{Tree} \\ \hline
\multirow{3}{*}{Importance} & \textit{Low}          & \multicolumn{1}{c|}{17} & \multicolumn{1}{c|}{17} & \multicolumn{1}{c|}{19}     & 20    \\ \cline{2-6} 
                            & \textit{Regular}      & \multicolumn{1}{c|}{14} & \multicolumn{1}{c|}{14} & \multicolumn{1}{c|}{16}     & 17    \\ \cline{2-6} 
                            & \textit{High}         & \multicolumn{1}{c|}{5}  & \multicolumn{1}{c|}{4}  & \multicolumn{1}{c|}{11}     & 10    \\ \hline
\multirow{3}{*}{Severity}   & \textit{Low}          & \multicolumn{1}{c|}{15} & \multicolumn{1}{c|}{14} & \multicolumn{1}{c|}{17}     & 16    \\ \cline{2-6} 
                            & \textit{Regular}      & \multicolumn{1}{c|}{11} & \multicolumn{1}{c|}{11} & \multicolumn{1}{c|}{3}     & 12    \\ \cline{2-6} 
                            & \textit{High}         & \multicolumn{1}{c|}{4}  & \multicolumn{1}{c|}{3}  & \multicolumn{1}{c|}{6}      & 5     \\ \hline
\multirow{3}{*}{Centrality} & \textit{Low}          & \multicolumn{1}{c|}{15} & \multicolumn{1}{c|}{15} & \multicolumn{1}{c|}{17}     & 16    \\ \cline{2-6} 
                            & \textit{Regular}      & \multicolumn{1}{c|}{12} & \multicolumn{1}{c|}{11} & \multicolumn{1}{c|}{14}     & 13    \\ \cline{2-6} 
                            & \textit{High}         & \multicolumn{1}{c|}{4}  & \multicolumn{1}{c|}{13} & \multicolumn{1}{c|}{8}      & 7     \\ \hline
\end{tabular}
\end{table}

Taken together, NVI, TTRG, TTPV, and VIB provide complementary evidence on policy behavior. Low values of TTPV and VIB indicate a well-managed backlog and fast remediation, but they do not necessarily translate into a lower goal achievement rate for the attacker, since the effectiveness of a patching strategy also depends on how well it aligns with the attacker's specific objective. Therefore, assessing policy effectiveness requires jointly considering NVI, TTRG, TTPV, and VIB together with the goal achievement rate and the nature of the attacker's objective.
\section{Threats to validity}
\label{sec:threats}

We here analyze threats to the validity of our study, and the methodological countermeasures to mitigate adverse effects on our findings.

As for \emph{external validity}, the main threat is related to the realism of the modeled attacker. To mitigate this, we designed \toolname{} to be highly configurable, allowing it to strictly align with data from Cyber Threat Intelligence (CTI), which provides information about exploited vulnerabilities and attack techniques adopted by real-world attackers. This approach leverages the best knowledge available from expert-analyzed CTI about APTs, which prioritizes well-documented APTs over mere chronological recency; specifically, we focused on vulnerabilities from 2020, as they coincide with the peak activity period of the public-documented APTs. 
A second fundamental challenge relates to the representativeness of the network scenarios. Public CTI withholds sensitive details (e.g., network topology, node configurations) to preserve confidentiality of affected organizations. For example, about the Equifax breach mentioned before, only the exploited vulnerabilities are publicly documented, but the underlying network structure and configuration remain undisclosed. Therefore, in our experiments, we considered typical enterprise network architectures. Furthermore, we mitigated this threat by designing \toolname{} to be fully configurable by practitioners to reflect their own organizations.

As for \emph{internal validity}, it concerns whether the underlying design choices are sound and complete. In particular, we adopted RL to orchestrate the ordering and timing of actions by attackers, with the action space grounded in MITRE ATT\&CK. Similarly, the defender's action space is currently focused on well-known policies for vulnerability management to orchestrate the actions of defenders. A direct consequence of this modelling choice is that the RL attacker is explicitly rewarded for targeting critical assets. While our abstraction might not capture every possible real-world attack or defense vector, we preserved this alignment because it realistically reflects goal-oriented APTs targeting an organization's pivotal assets. 
A second potential threat to internal validity relates to parameter sensitivity. To mitigate this, we derived the RL hyperparameters from previous research in the field. Furthermore, we conducted evaluations across different defender effort levels to ensure that the relative performance of the prioritization remains consistent across different operational constraints. Stochasticity in RL algorithms poses a threat to internal validity, as the attacker's converged policy might depend on the random seed initialization. To mitigate this, the reported likelihood of success of the attack campaigns was measured over multiple independent simulation runs, ensuring that our evaluation of the defender policies is not biased by a single, potentially anomalous, RL training outcome. 
We designed \toolname{} to be extendable in order to enable future research, as discussed in the next section. 

\section{Conclusion}
\label{sec:conclusion}

This work presents \toolname{}, a novel simulation tool for the quantitative evaluation of vulnerability management strategies. It captures the dynamics of complex attack campaigns in the context of continuous vulnerability management. By modeling different vulnerability management policies and the constraints imposed by a limited budget, it provides actionable insights into their effectiveness at mitigating APTs. 
Our experiments show highlight the need for vulnerability management strategies tailored to the organizational context, accounting for adversarial behavior, network topology, and asset criticality.

\toolname{} is a useful resource for security analysts, which can configure their own specific network environments and threat profiles. 
New APT profiles, network topologies, and vulnerabilities can be customized through configuration files, e.g., by specifying attack actions and targeted products for a specific APT, or by populating the network topology from existing IT asset management tools, such as CMDB systems. 

The tool is also a resource for further research on vulnerability management policies. The emergence of LLM-based vulnerability discovery will increase the volume of exploited vulnerabilities, which will require more accurate prioritization based on the organizational and threat context. New vulnerability management policies, such as policies based on the semantic analysis of the context and of the vulnerabilities through LLMs, are open research opportunities in the field. \toolname{} can be easily extended with new prioritization policies to be evaluated.

More broadly, \toolname{} is open to further extensions to model additional aspects that interact with vulnerability management, including SOC teams and intrusion detection processes, honeypots, and moving target defense strategies. 
The agent-based architecture of \toolname{} readily supports the integration of additional agents, such as a threat hunter agent performing periodic scans, a moving target defense agent relocating critical assets, or a topology update agent, which simulates network dynamism over time. These extensions do not require structural changes to the framework. We devised \toolname{} as a basis to enable further research in vulnerability management. We release the full source code and experimental setup for future research.

\bibliographystyle{ACM-Reference-Format}
\bibliography{references}

@article{Singh2019,
  author    = {Saurabh Singh and Pradip Kumar Sharma and Seo Yeon Moon and Daesung Moon and Jong Hyuk Park},
  title     = {{A comprehensive study on APT attacks and countermeasures for future networks and communications: Challenges and solutions}},
  journal   = {J. of Supercomp.},
  volume    = {75},
  number    = {8},
  year      = {2019}
}

@article{nurse2025patch,
  title={To Patch or Not to Patch: Motivations, Challenges, and Implications for Cybersecurity},
  author={Nurse, Jason RC},
  journal={arXiv preprint arXiv:2502.17703},
  year={2025}
}

@article{wang2018cybersecurity,
  title={{Cybersecurity incident handling: A case study of the Equifax data breach}},
  author={Wang, Ping and Johnson, Christopher},
  journal={Issues in Inf. Systems},
  volume={19},
  number={3},
  year={2018}
}

@inproceedings{feng2022defense,
  title={{Defense-in-depth security strategy in LOG4J vulnerability analysis}},
  author={Feng, Sylvia and Lubis, Muharman},
  booktitle={Intl. Conf. Adv. in Data Science, E-learning and Inf. Sys. (ICADEIS)},
  year={2022}
}

@misc{CVEorg,
  author       = {{MITRE}},
  title        = {{CVE - Common Vulnerabilities and Exposures}},
  year         = {2026},
  howpublished = {\url{https://www.cve.org}}
}

@misc{CISAKEV,
  author       = {{CISA}},
  title        = {{Known Exploited Vulnerabilities Catalog (KEV)}},
  year         = {2026},
  url          = {https://www.cisa.gov/known-exploited-vulnerabilities-catalog}
}

@misc{Gymnasium,
  author       = {{Farama Foundation}},
  title        = {{Gymnasium: An API standard for reinforcement learning with a diverse collection of reference environments}},
  year         = {2026},
  url          = {https://gymnasium.farama.org/}
}

@misc{NetworkX,
  author       = {{NetworkX Developers}},
  title        = {{NetworkX}: Software for Complex Networks},
  year         = {2024},
  url          = {https://networkx.org/}
}

@misc{NAO2017,
  author    = {{UK National Audit Office}},
  title     = {{Investigation: WannaCry cyber attack and the NHS}},
  year      = {2017},
  url       = {https://www.nao.org.uk/reports/investigation-wannacry-cyber-attack-and-the-nhs/}
}

@misc{CVSS,
  author       = {FIRST},
  title        = {{Common Vulnerability Scoring System (CVSS)}},
  year         = {2026},
  url = {https://www.first.org/cvss/}
}

@misc{EPSS,
  author       = {{FIRST}},
  title        = {{Exploit Prediction Scoring System (EPSS)}},
  year         = {2025},
  url          = {https://www.first.org/epss/}
}

@misc{CISA_SSV,
  author       = {CISA},
  title        = {{Stakeholder-Specific Vulnerability Categorization (SSVC)}},
  year         = {2026},
  url          = {https://www.cisa.gov/stakeholder-specific-vulnerability-categorization-ssvc}
}

@techreport{NIST_CSWP_29,
  author      = {NIST},
  title       = {The {NIST Cybersecurity Framework (CSF)} 2.0},
  institution = {NIST},
  number      = {CSWP 29},
  year        = {2024}
}

@misc{mitre_attack_ta0001,
  author       = {{MITRE}},
  title        = {Initial Access},
  howpublished = {\url{https://attack.mitre.org/tactics/TA0001/}}
}

@article{ADAWADKAR2022,
author = {Amrin Maria Khan Adawadkar and Nilima Kulkarni},
title = {{Cyber-security and reinforcement learning - A brief survey}},
journal = {Eng. Applications of Artificial Intelligence},
volume = {114},
year = {2022}
}

@article{farris2018vulcon,
  title={{VULCON: A System for Vulnerability Prioritization, Mitigation, and Management}},
  author={Farris, Katheryn A and Shah, Ankit and Cybenko, George and Ganesan, Rajesh and Jajodia, Sushil},
  journal={ACM Trans. on Priv. and Sec.},
  volume={21},
  number={4},
  year={2018}
}

@article{puterman1990markov,
  title={Markov decision processes},
  author={Puterman, Martin L},
  journal={Handbooks in operations research and management science},
  volume={2},
  pages={331--434},
  year={1990},
  publisher={Elsevier}
}

@book{sigaud2013markov,
  title={Markov decision processes in artificial intelligence},
  author={Sigaud, Olivier and Buffet, Olivier},
  year={2013},
  publisher={John Wiley \& Sons}
}

@article{watkins1992q,
  title={{Q-learning}},
  author={Watkins, Christopher JCH and Dayan, Peter},
  journal={Machine learning},
  volume={8},
  pages={279--292},
  year={1992},
  publisher={Springer}
}

@inproceedings{fan2020theoretical,
  title={{A theoretical analysis of deep Q-learning}},
  author={Fan, Jianqing and Wang, Zhaoran and Xie, Yuchen and Yang, Zhuoran},
  booktitle={Learning for Dynamics and Control},
  year={2020}
}

@techreport{NIST80040r4,
  author       = {Souppaya, Murugiah and Scarfone, Karen},
  title        = {{Guide to Enterprise Patch Management Planning: Preventive Maintenance for Technology}},
  institution  = {NIST},
  number       = {SP 800-40r4},
  year         = {2022}
}

@misc{NCSC2023,
  author       = {{National Cyber Security Centre}},
  title        = {Verify and regularly review your vulnerability management process},
  year         = {2023},
  howpublished = {\url{https://www.ncsc.gov.uk/collection/vulnerability-management/guidance/verify-review-process}}
}

@inproceedings{younis2015comparing,
  title={{Comparing and evaluating CVSS base metrics and Microsoft rating system}},
  author={Younis, Awad A and Malaiya, Yashwant K},
  booktitle={IEEE Intl. Conf. on Software Quality, Reliability and Security (QRS)},
  year={2015}
}

@article{wiebe2023learning,
  title={Learning cyber defence tactics from scratch with multi-agent reinforcement learning},
  author={Wiebe, Jacob and Mallah, Ranwa Al and Li, Li},
  journal={arXiv preprint arXiv:2310.05939},
  year={2023}
}

@techreport{recordedfuture2024vulns,
  title={{Patterns and Targets for Ransomware Exploitation of Vulnerabilities: 2017–2023}},
  author={{Insikt Group}},
  institution={Recorded Future},
  year={2024},
  note={\url{https://www.recordedfuture.com/research/patterns-targets-ransomware-exploitation-vulnerabilities-2017-2023}}
}

@techreport{csw2024ransomware,
  title={{Spotlight Report 2024: Ransomware Through the Lens of Threat and Vulnerability Management}},
  author={{Cyber Security Works}},
  year={2024},


}

@techreport{bitsight_critical_updates,
  title = {{A Growing Risk Ignored: Critical Updates}},
  author = {{BitSight Technologies, Inc.}},
  institution = {BitSight},
  note = {\url{https://info.bitsight.com/bitsight-insights-a-growing-risk-ignored-critical-updates}},
  year = {2017},
}

@misc{garrity_insights_2023,
  author       = {Patrick Garrity},
  title        = {Insights Into Vulnerability Management},
  year         = {2023},
  howpublished = {\url{https://nucleussec.com/blog/insights-into-vulnerability-management-v1/}}
}

@misc{islam_cvss_deception_2024,
  author       = {Syed Islam and Ankur Sand},
  title        = {The {CVSS} Deception: How We've Been Misled on Vulnerability Severity},
  year         = {2024},
  howpublished = {\url{https://i.blackhat.com/EU-24/Presentations/EU-24-Islam-The-CVSS-Deception.pdf}}
}

@techreport{cisco_p2p_vol9_2023,
  author      = {{Cisco} and {Cyentia Inst.}},
  title       = {Prioritization to Prediction, Vol. 9: Role of the Known Exploited Vulnerability Catalog in Risk-Based Vulnerability Management},
  year        = {2023},
  institution = {Cisco Systems}
}

@techreport{edwards_bitsight_kev_2024,
  author      = {Ben Edwards},
  title       = {A Global View of the {CISA KEV} Catalog: Prevalence and Remediation},
  institution = {Bitsight Technologies},
  year        = {2024}
}

@misc{redhat_cvss_not_risk_2019,
  author       = {{Red Hat}},
  title        = {{Why CVSS Does Not Equal Risk: How to Think About Risk in Your Environment}},
  year         = {2019},
  howpublished = {\url{https://www.redhat.com/en/blog/why-cvss-does-not-equal-risk-how-think-about-risk-your-environment}}
}

@misc{fernao_cvss_misconceptions_2024,
  author       = {{Fernao Group}},
  title        = {{CVSS}: The 5 Biggest Misconceptions in Risk Assessment},
  year         = {2024},
  howpublished = {\url{https://www.fernao.com/de/blog/cvss-the-5-biggest-misconceptions-in-risk-assessment}}
}

@misc{shick_towards_cvss_2018,
  author       = {Deana Shick},
  title        = {Towards Improving {CVSS}},
  year         = {2018},
  howpublished = {\url{https://insights.sei.cmu.edu/blog/towards-improving-cvss/}}
}

@misc{bansal2020zoning,
  author    = {Aman Bansal},
  title     = {Security Zoning in Network Architecture},
  year      = {2020},
  url       = {https://medium.com/@aman.bansal93/security-zoning-in-network-architecture-ff7693b91556}
}

@misc{metasploit,
  title        = {{Metasploit - Penetration Testing Software}},
  author       = {{Rapid7}},
  howpublished = {\url{https://www.metasploit.com}},
  year         = {2026}
}

@misc{projectdiscovery_nuclei,
  title        = {Nuclei: A fast and customisable vulnerability scanner},
  author       = {{ProjectDiscovery}},
  howpublished = {\url{https://docs.projectdiscovery.io/opensource/nuclei/}},
  year         = {2026}
}

@INPROCEEDINGS{cti-hal,
  author={Della Penna, Sofia and Natella, Roberto and Orbinato, Vittorio and Parracino, Lorenzo and Pianese, Luciano},
  booktitle={8th Workshop on Attackers and Cyber-Crime Operations (WACCO)}, 
  title={{CTI-HAL: A Human-Annotated Dataset for Cyber Threat Intelligence Analysis}}, 
  year={2025}
}

@misc{nist_nvd,
  author       = {{National Institute of Standards and Technology (NIST)}},
  title        = {{National Vulnerability Database (NVD)}},
  howpublished = {\url{https://nvd.nist.gov}},
  year         = {2026}
}

@misc{mitre-apt41,
  author       = {MITRE},
  title        = {{APT41}},
  howpublished = {\url{https://attack.mitre.org/groups/G0096/}},
  year         = {2026}
}

@misc{mitre-apt28,
  author       = {{MITRE}},
  title        = {{APT28}},
  year         = {2026},
  howpublished          = {\url{https://attack.mitre.org/groups/G0007/}},
}

@misc{ArcticWolf2025_ThreatReport,
  author       = {{Arctic Wolf}},
  title        = {2025 Threat Report},
  year         = {2025},
  url          = {https://www.arcticwolf.com/resource/arctic-wolf-threat-report-2025-lp/arctic-wolf-threat-report-2025}
}

@misc{Coalition2025_CyberThreatIndex,
  author       = {{Coalition, Inc.}},
  title        = {Cyber Threat Index 2025: Deciphering the Ransomware Playbook},
  year         = {2025},
  url          = {https://web.coalitioninc.com/DLC-Cyber-Threat-Index-2025.html}
}

@misc{Sophos2025_ActiveAdversaryReport,
  author       = {John Shier and Angela Gunn and Hilary Wood},
  title        = {{Sophos Active Adversary Report}},
  year         = {2025},
  url          = {https://news.sophos.com/en-us/2025/04/02/2025-sophos-active-adversary-report/}
}

@misc{lockheed_cyber_kill_chain,
  title        = {{Cyber Kill Chain®}},
  author       = {{Lockheed Martin Corp.}},
  howpublished = {\url{https://www.lockheedmartin.com/en-us/capabilities/cyber/cyber-kill-chain.html}},
  year         = {2025}
}

@misc{mitre_attack,
  title        = {{MITRE ATT\&CK: Adversarial Tactics, Techniques, and Common Knowledge}},
  author       = {{MITRE Corp.}},
  howpublished = {\url{https://attack.mitre.org/}},
  year         = {2025}
}

@misc{purplesec_vulnerability,
  title        = {{What is Vulnerability Management? (The Definitive Guide)}},
  author       = {{PurpleSec}},
  howpublished = {\url{https://purplesec.us/learn/what-is-vulnerability-management/}},
  year         = {2024}
}

@misc{wiz_vulnerability,
  title        = {{What is Continuous Vulnerability Management?}},
  author       = {{Wiz, Inc.}},
  howpublished = {\url{https://www.wiz.io/academy/continuous-vulnerability-management}},
  year         = {2025}
}

@misc{sentinelone_vulnerability,
  title        = {{What is Continuous Vulnerability Management?}},
  author       = {{SentinelOne}},
  howpublished = {\url{https://www.sentinelone.com/cybersecurity-101/cybersecurity/what-is-continuous-vulnerability-management/}},
  year         = {2025}
}

@article{mnih2013playing,
  title={{Playing Atari with Deep Reinforcement Learning}},
  author={Mnih, Volodymyr and Kavukcuoglu, Koray and Silver, David and Graves, Alex and Antonoglou, Ioannis and Wierstra, Daan and Riedmiller, Martin},
  journal={arXiv preprint arXiv:1312.5602},
  year={2013}
}

@misc{sei2022epss,
  title={{Probably Don’t Rely on EPSS Yet}},
  author={Spring, Jonathan},
  howpublished = {\url{https://www.sei.cmu.edu/blog/probably-dont-rely-on-epss-yet/}},
  year={2022}
}

@misc{empirical2026capacity,
  title = {{Capacity is King}},
  author = {Roytman, Michael},
  howpublished = {\url{https://research.empiricalsecurity.com/research/capacity-is-king}},
  year = {2026}
}

@misc{google_apt,
  title = {{APT groups and threat actors}},
  author = {{Google LLC}},
  howpublished = {\url{https://cloud.google.com/security/resources/insights/apt-groups}},
  year = {2026}
}

@misc{microsoft_apt,
  title = {{What is an advanced persistent threat (APT)?}},
  author = {{Microsoft Corp.}},
  howpublished = {\url{https://www.microsoft.com/en-us/security/business/security-101/what-is-advanced-persistent-threat-apt}},
  year = {2026}
}

@misc{cveicu,
  title = {{CVE Publications by Year}},
  author = {Gamblin, Jerry},
  howpublished = {\url{https://cve.icu/}},
  year = {2026}
}

@misc{vulnerability_stats,
  title = {{2017 - 2026 Vulnerability Severity By Year}},
  author = {Ferguson, Dave},
  howpublished = {\url{https://securityvulnerability.io/}},
  year = {2026}
}

@misc{ibm_topology,
  title = {{Hierarchical Tree Topology}},
  author = {{IBM Corp.}},
  howpublished = {\url{https://www.ibm.com/docs/en/informix-servers/15.0.x?topic=systems-high-availability-clusters-in-hierarchical-tree-topology}},
  year = {2026}
}

@book{industrial_cybersec,
  title = {{Practical Industrial Cybersecurity: ICS, Industry 4.0, and IIoT}},
  author = {Brooks, Charles J. and Craig Jr., Philip A.},
  year = {2022},
  publisher = {Wiley}
}

@misc{oracle_dmz,
  title = {{Deploying Services Gatekeeper in a Demilitarized Zone}},
  author = {{Oracle Corp.}},
  url={https://docs.oracle.com/communications/E81149_01/doc.70/e95424/sgsec_dmz.htm#SGSEC215},
  year = {2018}
}

@inproceedings{ou2005mulval,
  title={{MulVAL: A logic-based network security analyzer}},
  author={Ou, Xinming and Govindavajhala, Sudhakar and Appel, Andrew W and others},
  booktitle={USENIX Security Symp.},
  year={2005}
}

@incollection{singhal2017security,
  title={Security risk analysis of enterprise networks using probabilistic attack graphs},
  author={Singhal, Anoop and Ou, Xinming},
  booktitle={Network security metrics},
  year={2017},
  publisher={Springer}
}

@article{munoz2017exact,
  title={Exact inference techniques for the analysis of Bayesian attack graphs},
  author={Mu{\~n}oz-Gonz{\'a}lez, Luis and Sgandurra, Daniele and Barr{\`e}re, Mart{\'\i}n and Lupu, Emil C},
  journal={IEEE Transactions on Dependable and Secure Computing},
  year={2017},
  publisher={IEEE}
}

@misc{cyberbattlesim,
  author = {{Microsoft Defender Research Team}},
  title = {{CyberBattleSim}},
  year = {2021},
  url={https://www.microsoft.com/en-us/research/project/cyberbattlesim/}
}

@misc{becker2024evaluationreinforcementlearningautonomous,
      title={Evaluation of Reinforcement Learning for Autonomous Penetration Testing using A3C, Q-learning and DQN}, 
      author={Norman Becker and Daniel Reti and Evridiki V. Ntagiou and Marcus Wallum and Hans D. Schotten},
      year={2024},
      eprint={2407.15656},
      archivePrefix={arXiv},
      primaryClass={cs.CR},
      url={https://arxiv.org/abs/2407.15656}, 
}

@misc{schwartz2019autonomouspenetrationtestingusing,
      title={Autonomous Penetration Testing using Reinforcement Learning}, 
      author={Jonathon Schwartz and Hanna Kurniawati},
      year={2019},
      eprint={1905.05965},
      archivePrefix={arXiv},
      primaryClass={cs.CR},
      url={https://arxiv.org/abs/1905.05965}, 
}

@misc{anthropic2026glasswing,
  author       = {{Anthropic}},
  title        = {Project {Glasswing}: {Securing} Critical Software for the {AI} Era},
  year         = {2026},
  month        = apr,
  howpublished = {\url{https://www.anthropic.com/glasswing}},
  note         = {Accessed: 2026-07-05}
}

@techreport{ponemon2019vulnresponse,
  author      = {{Ponemon Institute LLC}},
  title       = {Costs and Consequences of Gaps in Vulnerability Response},
  institution = {Ponemon Institute, sponsored by ServiceNow},
  year        = {2019},
  url         = {https://media.bitpipe.com/io_15x/io_152272/item_2184126/ponemon-state-of-vulnerability-response-.pdf}
}

@techreport{qualys2026backlog,
  author      = {Abbasi, Saeed and {Qualys Threat Research Unit (TRU)}},
  title       = {The Broken Physics of Remediation},
  institution = {Qualys},
  year        = {2026},
  url         = {https://cdn2.qualys.com/docs/mktg/qualys-tru-the-broken-physics-of-remediation.pdf}
}

@techreport{first2026forecast,
  author      = {{FIRST.org}},
  title       = {Mid-Year Vulnerability Forecast 2026},
  institution = {Forum of Incident Response and Security Teams (FIRST)},
  year        = {2026},
  url         = {https://www.first.org/newsroom/releases/20260615}
}

@inproceedings{10.1145/3776942.3777008,
author = {Song, Xining and Li, Yuting and Zhang, Decai},
title = {Design and Implementation of Building a Small and Medium sized Enterprise Office LAN},
year = {2025},
isbn = {9798400715839},
publisher = {Association for Computing Machinery},
address = {New York, NY, USA},
url = {https://doi.org/10.1145/3776942.3777008},
doi = {10.1145/3776942.3777008},
booktitle = {Proceedings of the 2025 11th Annual International Conference on Network and Information Systems for Computers},
pages = {221–225},
numpages = {5},
location = {
},
series = {ICNISC '25}
}

@inproceedings{10.1145/3773365.3773603,
author = {Liu, Caiyan and Shen, Wentao and Lyu, Wenling and Xu, Xiaojie and Ling, Xufeng},
title = {A Study on network architectures and security for small and medium-sized enterprises},
year = {2025},
isbn = {9798400718748},
publisher = {Association for Computing Machinery},
address = {New York, NY, USA},
url = {https://doi.org/10.1145/3773365.3773603},
doi = {10.1145/3773365.3773603},
booktitle = {Proceedings of the 2025 8th International Conference on Computer Information Science and Artificial Intelligence},
pages = {1514–1519},
numpages = {6},
location = {
},
series = {CISAI '25}
}

\end{document}